\documentclass[manuscript]{aastex}
\usepackage{rotating}

\begin{document}


\title{$Kepler$ Cycle 1 Observations of Low Mass Stars: New Eclipsing 
Binaries, Single Star Rotation Rates, and the Nature and Frequency of Starspots}

\author{T. E. Harrison, J. L. Coughlin, N. M. Ule}

\affil{Department of Astronomy, New Mexico State University, Box 30001, MSC 
4500, Las Cruces, NM 88003-8001}

\email{tharriso@nmsu.edu, jlcough@nmsu.edu, nmule@nmsu.edu}

\and

\author{M. L$\acute{o}$pez-Morales}

\affil{Institut de Ciencies de L'Espai (CSIC-IEEC), Campus UAB, Fac. Ciencies.
Torre C5 parell 2, 08193 Bellaterra, Barcelona, Spain}

\email{mlopez@ieec.uab.es}

\begin{abstract}
We have analyzed $Kepler$ light curves for 849 stars with T$_{\rm eff}$ 
$\leq$ 5200 K from our Cycle 1 Guest Observer program. We identify six new eclipsing 
binaries, one of which has an orbital period of 29.91 d, and two of which are 
probably W UMa variables. In addition, we identify a candidate ``warm Jupiter''
exoplanet. We further
examine a subset of 670 sources for variability. Of these objects, 265 stars clearly
show periodic variability that we assign to rotation of the low-mass star. At
the photometric precision level provided by $Kepler$, 251 of
our objects showed no evidence for variability. We were unable to determine
periods for 154 variable objects. We find that 79\% of stars with 
T$_{\rm eff}$ $\leq$ 5200 K are variable. The rotation periods we derive for the 
periodic variables span the range 0.31 $\leq$ P$_{\rm rot}$ $\leq$ 126.5 d. 
A considerable number of stars with rotation periods similar to the solar value show 
activity levels that are 100 times higher than the Sun. This is consistent with 
results for solar-like field stars. As has been found in previous studies, stars with 
shorter rotation periods generally exhibit larger modulations. This trend flattens 
beyond P$_{\rm rot}$ = 25 d, demonstrating that even long period binaries may still 
have components with high levels of activity and investigating whether
the masses and radii of the stellar components in these systems are consistent
with stellar models could remain problematic. Surprisingly, our modeling of the light 
curves suggests that the active regions on these cool stars are either preferentially 
located near the rotational poles, or that there are two spot groups located at lower 
latitudes, but in opposing hemispheres.
\end{abstract}

\noindent
{\it Key words:} stars: low-mass --- stars: late-type --- binaries: eclipsing ---
stars: spots 

\section{Introduction}

The $Kepler$ mission was designed to discover and characterize transiting
exoplanetary systems (Borucki et al. 2010), and it has been quite successful
with more than 1200 candidate systems already identified (Borucki et al. 2011).
Perhaps equally exciting to the discovery and observation of exoplanets is
the impact of high precision, long-term photometric observations on our understanding
of ordinary
stars. These new data are providing a wealth of information on the
asteroseismology of stars similar to the Sun (e.g., Verner et al. 2011), as well 
as those objects in the classical regions of the instability strip (e.g.,
Benko et al. 2010). 

It is also possible to use $Kepler$ to explore important, outstanding issues that
remain for low-mass stars. One of the most important of these is that some of the
measured fundamental parameters of low-mass stars (masses, radii, and T$_{\rm eff}$)
appear to be in conflict with values predicted by models. For example, analysis by 
L$\acute{o}$pez-Morales (2007) shows that the observed radii of low-mass stars are 10 
to 20\% larger than predicted by the stellar models of Baraffe et al. (1998). 
L$\acute{o}$pez-Morales found that there was a clear correlation between the
activity levels of short period (P$_{\rm orb}$ $\sim$ 3 d) binaries and the 
discrepancy in the radii of the stellar components. Morales et
al. (2008, 2010) suggest that either the models are flawed, or that differences
in metallicity, magnetic activity, or the presence/distribution of star spots
make such comparisons moot. If the larger radii observed for the components in short 
period binaries is indeed due to enhanced magnetic activity as a result of tidal 
locking, then it is suspected that the low mass stars in binaries with orbital 
periods longer than $\sim$ 10 d should have radii in agreement with models. 
The main issue that existed before the launch of the $Kepler$ mission was the lack of 
long period (P$_{\rm orb}$ $>$ 10 d), low-mass eclipsing binaries.
The samples used to arrive at the results above relied on a handful of
shorter period eclipsing binaries. It is well known (Bopp 1987, Radick et al.
1987) that rapidly rotating low-mass stars are intrinsically more active, and 
thus if the components in short period eclipsing binaries are spun-up by
gravitational interactions (Zahn 1977, 1994), then their value for comparison to 
more slowly rotating isolated field dwarfs is diminished.

It would be extremely useful to discover additional low-mass eclipsing binaries
with longer periods to investigate whether these more slowly rotating objects
might be used to constrain stellar models. This was the genesis for our
Cycle 1 and 3 $Kepler$ programs that proposed to obtain light curves for 
2400 cool (T$_{\rm eff}$ $\leq$ 5200 K) stars to search for previously unknown 
long period, low-mass eclipsing binaries (LMEBs). As we discuss below, we have 
been successful in identifying an LMEB that has a period of 29.91 d, along 
with five shorter period LMEBs in our Cycle 1 data. This success rate is in line 
with our assumption that the mean binary star fraction is 35\% for K and M dwarfs
(Mayor et al. 1992, Leinert et al. 1997, and Fischer \& Marcy 1992).

Along with the discovery of new LMEBs, the $Kepler$ light curves
from our program objects provides an unprecedented data set to investigate 
the variability of an unbiased sample of low-mass main sequence stars on 
timescales of $\sim$ 90 d. For example, we can use the modulation of the light 
curves induced by starspots to determine the rotation rates for these stars, and 
compare them with predictions for the angular momentum evolution of such objects 
(e.g., Krishnamurthi et al.  1997). It is also possible to investigate the nature
of stellar activity for low-mass stars to levels not previously attainable using
ground-based photometry. With a large sample of objects with continuous light curves
spanning 90 (or more) days, we can probe the longevity of the active regions on these 
low-mass stars, and attempt to unravel the latitudinal
(and/or longitudinal) distribution of starspots on these objects.

In the following we discuss the selection of our targets and the reduction
of the Cycle 1 $Kepler$ light curves in section 2, discuss
the discovery of the new LMEBs and derivation of the stellar characteristics
in section 3, and discuss our conclusions in section 4.

\section{Observations}

Our sample of objects for observation in Cycle 1 (which ran from 2009 June 20,
until 2010 June 23) was selected using the stellar parameters available in the
{\it Kepler Input Catalog} (the ``KIC'', Brown et al. 2011). With multi-band
photometry ($griz$, ``DDO51'', and $JHK$ from 2MASS), Brown et al. were able to
assign values of T$_{\rm eff}$, log($g$), A$_{\rm V}$, and estimate the
metallicity for more than 4 million objects in the $Kepler$ field-of-view 
(note: see the discussion in Brown et al. for important caveats on each
of these derived parameters). From the KIC for Cycle 1, we constructed a list of 
all of the stars that had T$_{\rm eff}$ $\leq$ 5200 K,
$r$ $\leq $ 17.0, log($g$) $\geq$ 4.0, A$_{\rm V}$ $\leq$ 0.5, and that 
were not already designated as $Kepler$ program stars. In addition, we only 
selected targets that had no neighboring
objects within 6" of our sources. There were 1239 objects that met these
criteria. Dropping the faintest 39 sources, we ended
up with a list of 1200 targets. The brightnesses of our observed targets 
spanned the range 15.16 $\leq$ $r$ $\leq$ 16.97.

Our proposal sought to obtain ``long cadence'' (30 min) light curves spanning 90 
days (a single $Kepler$ ``Quarter''\footnotemark[1]\footnotetext[1]{See http://archive.stsci.edu/kepler/seasonstable.html for the start dates of each quarter.}) for
1200 late-type dwarfs. Unfortunately, due to a mission programming error, only 
849 unique targets were observed during Cycle 1. This did mean, however, that 260
targets were accidentally re-observed in a second quarter. In nearly all these cases
we obtained two quarters of data that {\it spanned} three quarters, skipping the 
middle quarter (e.g., quarter 2 + 4, or quarter 3 + 5). This separation allowed 
us to examine the light curves for coherent variations on timescales of up to 270 
days, and is especially useful in searching for very slowly rotating stars.

\subsection{Extraction of the Light Curves}

The ``Pre-search Data Conditioned'' (PDC) light curves released to Guest 
Observers (GOs) have been processed through
the $Kepler$ pipeline (see Jenkins et al. 2010). After careful examination,
the process used by the $Kepler$ team appears to introduce artifacts, and/or 
remove valid photometric variations. Part of the reason for this is that the
drift for some stars during a quarter was large enough that significant flux 
was lost out of the optimal photometric aperture used by the $Kepler$ pipeline 
to produce the light curves. While the fine guidance sensors on $Kepler$ 
generally keep the pointing to within $\pm$ 0.05 pixels, the effects of 
differential velocity aberration as the 
telescope orbits the Sun introduces an annual motion of 6" on the $Kepler$ 
field-of-view\footnotemark[2]\footnotetext[2]{See the Kepler Instrument Handbook 
at http://archive.stsci.edu/kepler/manuals/KSCI-19033-001.pdf for more details}. 
Subsequent polynomial fits to 
remove such trends appears to both introduce, and eliminate, photometric 
variations. Since it is impossible for a GO to reconstruct the process for any 
of the program targets, we decided to build our own processing pipeline that 
starts with the complete set of calibrated, sky-subtracted postage stamp images 
released to the GOs in early 2011.

The default photometric aperture assigned to each target depends on the
source brightness, and usually spans several pixels. This central aperture is
then surrounded by a ``sky'' region. For the faint stars of our program, the 
resulting postage stamp images were usually 3 $\times$ 5 or 4 $\times$ 6 pixels 
in size. Preliminary investigations showed, however, that a significant portion 
of the stellar flux was not contained within the pre-defined apertures for our 
targets. Thus, some of the stellar flux could often be detected in the ``sky'' 
pixels that surrounded every target's aperture. This is partially due to the
large point spread function of the defocussed images produced by $Kepler$, but 
mostly due to positional drift over the quarter. The result was that the PDC 
light curves have systematic variations due to improper ``sky'' removal, and too 
small of a photometric aperture.

As a result, we decided to sum-up the flux in {\it all} of the pixels in each
postage stamp image to get a better handle on the light curve of a source.
We found this process alone dramatically reduced the presence of the large scale 
systematics noted above. The main drawback to this method is that flux from 
nearby stars can often contaminate the resulting light curve. In the 
cases where this background severely contaminates the light curve of our program
object, we define those light curves as suffering from ``third light'' 
contamination. As noted 
below, many times such contamination can be non-varying, and be accounted for 
in extracting the target light curve. In other cases, it is impossible to
determine which object in the FOV is the source of the variations. To better
avoid such issues we would 
recommend that future observers of faint targets increase the nearest neighbor 
radius to $>$ 12" if there are any nearby stars that might provide sufficient
contaminating flux (i.e., stars as bright, or brighter than the target). The 
light 
curves generated by the summing process are then 
(only) used to determine the mean flux value for that source during the quarter.

After the flux summing process, the systematics in the light curve due to 
positional drift are greatly reduced, but pixel-to-pixel sensitivity, both in 
spectral response,
as well as quantum efficiency, and intrapixel variations, can produce significant
systematics. To account for these, we performed a Principal Component Analysis
(PCA; Murtagh \& Heck 1987) on the ensemble of post stamp images for each source.
The PCA process correlates the variation of the flux in each pixel with the
fluxes in all of the other pixels in the image. We then subtract the most 
significant principal component from the light curve and scale it using the mean
of the summed light curve to conserve flux. This latter step is done so that we 
can obtain light curve amplitudes that are calibrated to the $Kepler$ system. In 
this way, the effects noted above are largely accounted for, and the result is a 
relatively clean light curve. 

To demonstrate the process we present light curves for three targets in 
Fig. 1. In the top panels of this figure are the PDC ``calibrated''
pipeline processed $Kepler$ light curves as delivered to GOs. It is clear from 
the pipeline processed
light curve that Kepler ID\#8016381 (Fig. 1a) shows periodic variations, but 
they are somewhat irregular. The summed and PCA-processed light curves are 
much more symmetric. In the case of K8016381, these latter two light curves are 
identical. This result is due to the fact that the centroid of this star 
only moved a tiny fraction of a pixel over the 90 d quarter. In contrast to
K8016381 is the light curve for K5428432 (Fig. 1b). The PDC pipeline processed light 
curve suggests a possible slow, periodic variation with a significant amplitude. 
Our PCA processing reveals what appears to be a much more rapid oscillation of
lower amplitude. As shown in the centroid plot, however, this star appeared to
have numerous jumps in its position, and nearly every one of these is imprinted
on its light curve. These glitches have a variety of origins, from focus changes,
to cosmic ray hits, all of which are discussed in Jenkins et al. (2010). In the 
following, all light curves from sources with large and choppy centroid
motion have been removed from our rotation/starspot analysis. Such light curves 
remain useful for discovering LMEBs, exoplanet transits, or for short period
asteroseismology investigations (see Garcia et al.  2011 for one possible method 
to correct these types of light curves).

As mentioned above, some of our objects suffer from third light contamination.
Such contamination is easily identified by examining the motion of the centroid.
As shown in Fig. 1c, if the motion of the centroid of the star reflects the
variability seen in the light curve, there is significant third light 
contamination.
The amount of such contamination varied from an insignificant level, up to
to the point where it was so dominant as to render the light curve useless.
Where we felt that a light curve was only marginally affected, it was 
included in the analysis program, otherwise it was discarded. There was
no defined rule for such decisions, given the great range in behavior exhibited
by our program objects but, generally, if the third light imparted centroid
motion on order of $\pm$ 0.01 pixels, it was deemed to be problematic.
Of the 849 light curves that comprised our Cycle 1 survey, 173 were dropped from
further analysis either due to third light contamination (33), or from multiple 
large, abrupt changes in the centroid motion (140). 

\subsection{Light Curve Classification}

The $Kepler$ PDC pipeline-processed light curves, even with the flaws noted above,
were sufficient to search for eclipsing binaries. From the 849 targets of our
sample we identified six eclipsing binaries (two of which are almost certainly
W UMa variables). These objects, along with
their derived parameters, are listed in Table 1. After discarding the unusable 
light curves, we proceeded to analyze the light curves to classify the objects
as periodic variables, probable periodic variables of long period, aperiodic 
variables, and objects showing no variability. We define periodic variables
as objects that display either two similar, symmetric maxima with one minimum, or 
two minima with one maximum in their light curves (which either spanned 90, 180, 
or 270 d). Long period variables were objects that clearly showed evidence for 
both a minimum and a maximum, but the light curves were of insufficient duration 
to reveal a second maximum (or minimum). Aperiodic variables were objects with
complex light curves that ranged from quasi-periodic, to chaotic. Non-variables
were obviously objects that showed no significant variability. We found that we
could generally detect symmetric oscillations that had amplitudes
of $\sim$ 10 counts peak-to-peak in light curves that have mean fluxes of
$\sim$ 2000 counts. Thus, non-variables are objects whose {\it periodic} 
variability has a total amplitude below $\sim$ 0.5\%. These objects could be 
aperiodic, but at this level, tiny centroid changes can impart aperiodic 
features in the light curves, making such classification impossible. Of the 670 
light curves analyzed for variability, we found 265 periodic variables 
(Table 2), 126 long period variables (Table 3), 28 aperiodic variables 
(Table 4), and 251 non-variables (Table 5).

\subsection{The Sample}

After classification, we can examine the types of stars found in each group.
A histogram of the temperature distribution of our entire sample of 849 objects 
can be found in the top-left panel of  Fig. 2. Our 
sample is dominated by early K dwarfs. Our hottest stars have T$_{\rm eff}$ = 
5200 (K0V), and our coolest objects have T$_{\rm eff}$ = 3500 K (M2V).
Histograms of the temperature distributions of our long period variables, and 
the non-variable objects, are shown in the top-right, and bottom-left hand panels,
respectively, of Fig. 2. The histogram
for the long period variables closely resembles that of the overall sample,
while the histogram for the non-variable objects is dominated by 
hotter stars. It is clear that nearly all stars with T$_{\rm eff}$ $\leq$ 4000 K 
are variables at the level detectable with $Kepler$. This is reflected in the
histogram for the sample of periodic variables (bottom-right panel of Fig. 2) 
that demonstrates periodicities are more likely to be found in cooler stars. 
Basri et al. (2011) have performed a statistical analysis of the entire sample of
150,000+ program stars observed during Q1, and found that 87\% of their
(6522) stars with temperatures below 4500 K were variable. We find that 79\% of 
our sample of (298) stars with T$_{\rm eff}$ $\leq$ 4500 K are intrinsic variables.

Since the stellar activity level
increases as one descends the main sequence, it is obvious that rotational
modulation of the light curve due to starspots is what is driving the majority
of the periodic
variations. We will discuss this assumption in the next section. To better
characterize the periodic sample, we present a color-color plot of these objects
in Fig. 3. Even with the caveats in the assignment of T$_{\rm eff}$ in the
KIC ($\pm$ 200 K, or roughly one spectral type) noted by Brown et al. (2011), our set 
of periodic variables clearly has the desired high temperature cutoff near a spectral 
type of K0V. An HR diagram (Fig. 4) of 
the periodic variable sample indicates that most of these objects are early to
mid-K dwarfs with distances near 1 kpc. The height above the galactic plane
of our periodic variable sample is shown in Fig. 5. The mean scale height of
our sample is 249.9 pc. This is consistent with an ``intermediate disk
population'' (e.g., Ng et al. 1997) with a mean age near 5.75 Gyr. Thus, this
group of periodic variables should be dominated by objects similar to the Sun in 
both age and metallicity.

\subsubsection{Determining Rotation Periods}

We will assume that the periodic modulations in the observed light curves of our
sample of late-type stars is due to rotation. The association of photometric 
modulations in late-type stars with starspots dates back to at least Kron (1947). 
Vaughan et al.  (1981) obtained long term Ca II H and K photometry of a sample of 
ninety one F to M stars that showed days, to month long variations, in nearly all 
of their stars. Twenty of these stars clearly showed evidence for rotational 
modulation with periods ranging from 2.5 to 54 days. They found that these rotation
periods were consistent with spectroscopic determinations of the 
($v$sin$i$) rotation rates for their stars, confirming rotation as the source of 
the photometric variability. These results demonstrate that, unlike the Sun,
large regions of
stellar surface activity can last for several rotation periods. Dorren \& Guinan
(1983) observed several of the stars from Vaughan et al. in both narrow-band
and intermediate-band filters to sample both the blue continuum, and H-alpha 
spectral regions. For their most variable object (HD149661, K0V), total 
variations of 4\% in the blue, and 2\% in the red continuum filters were observed. 
Interestingly, they found that both the H-alpha and Ca II emission were 
anticorrelated to the continuum fluxes. They conclude that the photometric
modulations were due to dark spots on these stars (all of which probably have 
significantly higher levels of chromospheric activity than seen on the Sun).

To determine the rotation rates for our set of periodic variables we have
used Period04 (Lenz \& Breger 2005) to perform Fourier analysis to identify the
dominant frequencies in each light curve. One of the nice features of Period04 is 
the ability to remove low frequency modulations of the light curves to produce 
``pre-whitened'' light curves to precisely identify the period of the larger
amplitude modulations. There are numerous examples in our periodic
variable data set where the dominant periodic amplitudes are slowly modulated by
a longer term variation that is intrinsic to the star. In some cases this appears to
be due to changing starspot size, while in others it is simply a 
global change in the continuum level. For most objects, such slow modulations were
fit with a single minimum or maximum of a sinusoidal variation that had a period
that was many times longer than the duration of the light curve. 

In addition, there are a number of objects that show more complex light curves,
with a second period that could be identified. We believe that for the majority of
objects, this is due to a separate starspot group that is present on the star. In some 
cases, the periods for these second modulations are identical
to the primary modulation period, but are simply offset in phase. In other
sources, however, the second period is different to the primary period. We
believe that the best explanation for such period differences is differential
rotation. 

\section{Results}

The main goal of our program was to identify new LMEBs with periods in excess
of 10 days so as to test whether rotational spin-up was the genesis for the
discrepancy between the observed radii of low-mass stars, and predictions from
stellar models. From our Cycle 1 data we identified six new LMEBs. Two
of these objects, K636722 and K8086234, are almost certainly W UMa variables, and
technically not LMEBs. One of these six, K6431670, has a period in excess of 10 
days (see Table 1). Given that before the launch of $Kepler$ the longest known 
period of an LMEB was 8.4 d (Devor et al., 2008), this new object certainly met 
the program goals. This finding, however, is overshadowed 
by the results from analysis of the $Kepler$ public release data from Q1. 
Coughlin et al. (2011, see also Pr$\check{s}$a et al. 2011) identified 231 
eclipsing binaries in this data set where the primary star is cooler than the
Sun. Twenty nine of those LMEBs have periods longer than 10 d. As the majority 
of those systems are several magnitudes brighter than K6431670, 
they are much better binaries for determining whether the components in long 
period LMEBs have radii in line with the predictions from models. We are 
currently obtaining radial velocity and multi-wavelength light curves for a
subset of these 231 objects to allow for further investigation of this issue.

As noted above, the six LMEBs we have discovered are within expectations for
our sample size.  Mayor et al. (1992) estimate a binary fraction of 45\% for
K dwarfs, while for M dwarfs the binary fraction estimate ranges from 25\%
(Leinert et al. 1997) to 42\% (Fischer \& Marcy 1992). In a complete,
and unbiased survey, Delfosse et al. (2004) found a binary fraction of 
26 $\pm$ 3\% for M dwarfs. Since our sample draws from a mixture of these
two spectral types, we would expect binary fractions of $\approx$ 35\%.
The orbital period distribution for late type stars is not very well
known, but Fischer \& Marcy (1992) found a peak centered at 9 yr for M dwarfs. 
Their
result resembled that for G stars, which have a Gaussian distribution centered 
at $\langle$P$_{\rm orb}$$\rangle$ = 173 $\pm$ 8.6 yr (Duquennoy \& Mayor 1991). 
Putting these data together, assuming $\langle$P$_{\rm orb}$$\rangle$ = 9 $\pm$ 
8.6 yr, and a main sequence radius relationship, we estimate that in a sample of 
849 stars, we would expect to find five low mass eclipsing binaries.

In addition to the six LMEBs detected in our survey, we also discovered a 
candidate exoplanet system: K5164255 ($r$ = 16.37, designated as KOI824.01 by 
the $Kepler$ team). This object is probably a ``warm'' ($\sim$ 650 K) Jupiter 
due to its moderate orbital period (P$_{\rm orb}$ = 15.4 d), and the fact that 
the host star primary is a K3V (T$_{\rm eff}$ = 
4829). Our modeling of the $Kepler$ light curve for this object, presented
in Fig. 6, using JKTEBOP (Southworth et al. 2004a,b) leads to the 
parameters for the host star and planet 
listed in Table 6. Given the large number of such objects detected by 
$Kepler$, this object is of little interest for further investigation due 
to the faintness of its host star, which is below the capabilities of 
current radial velocity studies.

\subsection{Rotation Rates of Low Mass Stars}

We have assumed that the periodic modulation seen in the light curves (e.g., Fig.
1) is the rotation period of those stars. In Table 5 we list the periods of the 
dominant modulation found from fitting the light curves using Period04. The 
derived periods range from 0.31 d to 126.5 d, with a mean of 32.12 d.
We present a plot of rotation period vs. T$_{\rm eff}$ in Fig. 7. The resulting
distribution is remarkably flat (the means in 200 K bins are also plotted).

As discussed earlier, rotation of a spotted star is the obvious interpretation for
the modulations we detect in the light curves of our late-type dwarfs. As shown by 
Christensen-Dalsgaard \& Frandsen (1983), solar-like 
oscillations in late-type dwarf stars have similar pulsation periods as the Sun 
($\sim$ 5 min), with amplitudes of a few parts-per-million. Such oscillations 
could never be seen in the $Kepler$ long duration light curves. Gilliland (2008) 
has shown that red giants have variability on these time scales, 
but those variations have a recognizable photometric signature. Basri 
et al. (2011) investigated this aperiodic signature to see if it is possible to 
select giants vs. dwarfs based upon their $Kepler$ light curves alone. They found 
that the selection process was robust, with only a few high gravity objects 
showing up in their analysis as red giants. Basri et al. suggest that it is 
possible that these few objects are actually red giants that were misclassified as 
dwarfs in the KIC. This process demonstrates that the gravities listed in the KIC 
are fairly reliable for late-type stars. 

The light curves of red giants can show quasi-periodic behavior that are similar
in nature to the oscillations seen in the Sun, except for their longer periods and 
much greater
amplitudes (see Huber et al. 2010). We attempted to fit periods to all objects whose 
light curves might be periodic. Only after we could not identify a period consistent
with the {\it entire} light curve were those objects re-classified as aperiodic. It 
is highly likely that {\it no} red giants remain in our ``rotation'' sample.

It is possible, however, that the period we determine for the rotation rate is an even
multiple, or fraction, of the true period. Shown in Fig. 8 is an example of
an object (K10200948) that has one of the more complicated, and rapidly evolving
light curves of any of the sources in our sample. In the Q3 data, analysis using
Period04 finds a period of 7.197 d. In the Q5 data, we find the dominant
period to be 14.45 d. Analysis of the combined light curves results in a best-fit
period of 14.006 d. This latter period is what is over-plotted on the light curves
shown in Fig. 8, and is what we assign as the rotation period for this object. 
Clearly, additional spots are present on K10200948 that are moving/evolving relative 
to the ``main spot group'' responsible for the coherent oscillation that retains the 
exact same phasing
over 270 d. Fortunately for this target we have two quarters of data that allows
us to isolate the ``correct'' period for this target. It is obvious, however, that
there could be a few cases where similar sets of starspot groups are found to be 
centered on opposite hemispheres so as to create confusion about the true period. It is 
also obvious that if differential rotation is present, as suggested by the changing
shapes of the maxima in the light curve of K10200948, the period we derive 
could be incorrect due to the slow migration of the main spot group. Since we have
no secondary information as to the inclination of the rotational axis of these
stars, it is impossible to know the exact latitude of these active regions, or whether 
the large spot groups are actually changing position with time, or are simply evolving
in size and/or shape. 

Another possible effect that could cause errant periods is evolution
of a starspot group on a timescale similar to the rotation period. If the 
stellar activity was somehow constrained to evolve at certain discrete 
longitudes, then the appearance, or disappearance, of spot groups with
lifetimes similar to the rotation period could lead to erroneous period 
determinations. This is most
true for the longest period systems. Instead of rotation, these light curves could 
be modeled by the repeated growth and decay of a fixed single spot group on a 
non-rotating star. Hopefully, nature is not this cruel, but since our knowledge of 
stellar activity cycles and the rotation rates of low-mass field stars is still 
primitive, it is impossible to rule out such behavior. 

\subsection{Amplitude of the Spot Modulation}

Tabulated in Table 5 are the amplitudes of the modulations for our sample of
periodic objects. The amplitude listed there is the {\it largest peak-to-peak}
variation seen in the light curve of that object. We plot the distribution of
these amplitudes with respect to temperature (as well as their means) in Fig. 9.
Surprisingly, this distribution is flat, with the early K dwarfs showing a greater
range in modulation than the M-type dwarfs. The temperature-dependent means (in 
200 K bins), however, are consistent with a single value (the sample mean was 
1.3\%).

The level of variation seen here is consistent with the observations of Dorren \& 
Guinan (1983), suggesting that our sample of rotating stars has a higher level of 
activity than exhibited by the Sun. However, it is important to realize that a nearly 
equal number of stars in our survey were found to exhibit no variations, though even
those stars could be more active than the Sun. Treating
the Sun as a variable star using the VIRGO\footnotemark[3]\footnotetext[3]{See http://www.ias.u-psud.fr/virgo/} instrument on SOHO, Lanza et al. (2004)
show that the white light (similar to $Kepler's$ filterless response 
function\footnotemark[4]\footnotetext[4]{http://keplergo.arc.nasa.gov/CalibrationResponse.shtml}, see below) variations during solar maximum are of order of 500 ppm (see also
Pagano et al. 2005), a factor of ten smaller than our detection limit of $\sim$ 0.5\%.

It is also interesting to examine the rotation period-activity relationship.
In Fig. 10 we plot the amplitude of the variations vs. the rotation period for
our sample of periodic variables. It has been well established that younger stars 
rotate more quickly, and that these objects display a higher level of activity (c.f.,
Radick et al. 1987). This trend is observed in our sample, where the most rapid
rotators generally show larger photometric modulations. At periods longer than
$\sim$ 25 d, the trend flattens dramatically. We discuss the implications of this
result in the next section.

\subsection{Spot Modeling of the Light Curves}

Without additional observations, it is difficult to derive the inclination angles
of the rotational axes for any of the stars in our sample. Statistically, the
mean inclination angle for a random sampling of rotating stars is $i$ = 
57$^{\circ}$. There are two light curve morphologies that can be generated using
a {\it single spot}: continuously variable (sinusoidal), and flat maxima. Continuously
variable light curves occur when the starspot is ``circumpolar'' in the sense that
it is always in view for the observer. Flat-maxima light curves occur when the spot 
is out of view for some fraction of the rotation period. It is interesting to examine 
what the break-down into these two categories might imply for the latitudinal
distribution of spots.

We have examined the light curves of all our rotation targets, and have classified 
them into the following groups: flat-maxima (90 objects), continuously variable (119), 
flat-minima (40), and complex (16). A flat maximum light curve simply has maxima
that are broader than the minima seen in that light curve. Flat minima light curves
are the opposite to flat maxima. Continuously variable light curves have minima and 
maxima that have nearly identical shapes. Complex light curves have such dramatically 
changing spot groups that it is not possible to identify a consistent portion of the 
light curve that allows classification into the one of the other three categories. 
Complex and flat minima light curves cannot be explained using a single spot. A light 
curve with a flat minimum suggests that a second spot, with similar properties, rotates 
into view as the first spot is passing out of view. The flat-minima light curves in 
our sample always have non-flat maxima. 

The fact that 57\% of our rotation sample have continuously variable light curves
suggests that for the majority of our objects the sum of the inclination angle ($i$)
and the starspot co-latitude ($b$, formally defined below) is less than 90$^{\circ}$. 
This relationship
simply states that the starspot remains in view at all times (especially given that a 
starspot must subtend an appreciable angle to be detectable, see below). If we
assume a random set of rotational inclination angles (0.0 $\leq$ $i$ $\leq$ 90.0), and 
a random value for the starspot's co-latitude (0.0 $\leq$ $b$ $\leq$ 180.0), we find 
that statistically, only 21\% of our stars should have continuously variable light 
curves. Assuming single, dominant spot groups, this result would strongly argue for 
``polar'' spots.  The only alternative to this 
conclusion is that many of our stars have two spot groups in diametrically 
opposite hemispheres with similar enough properties to create continuously variable 
light curves. The fact that 40 of our stars have flat minima, an indicator for two
very similar spot groups, indicates that this latter conclusion may have some
validity. 

We can attempt to model the light curves of our rotating stars to investigate 
starspot sizes and distributions, but before we do so, it is important to establish 
what effect the broad bandpass of 
the $Kepler$ mission has on the detection of cool spots. 
Basri et al.  (2010) have investigated the variability of solar-like stars with an 
emphasis on comparison to the Sun. They found that the $g+r$ light curves from 
the VIRGO
instrument on SOHO essentially reproduces the broad $Kepler$ bandpass.
Thus, the minima in the $Kepler$ light curves of cool stars are due to dark spots on 
the surface, and thus we are not seeing maxima due to bright faculae (that actually cover a larger 
fraction of the solar photosphere than spots). As discussed in Knaack et al.
(2001), faculae have a higher contrast near the limb of the Sun in the visual
bandpass. In contrast, sunspots have a higher contrast when located near the center 
of the solar disk. Knaack et al. investigated whether viewing the Sun from higher 
inclination angles would result in a greater photometric variation, making the 
Sun more consistent with the larger variability observed for field stars {\it like} 
the Sun. They found that changing the Sun's inclination has only a modest ($\sim$ 
+6\%) effect on the solar irradiance. Thus, the $Kepler$ light curves should be
useful for constraining the properties of starspots for our sample of randomly
inclined, rotating stars.

To investigate the starspot parameters for the objects in our rotational sample we 
have modeled the $Kepler$ light curves using PHOEBE (Pr$\check{s}$a \& Zwitter 2005),
a graphical interface to the Wilson-Divinney binary star light curve modeling code 
(see
Kallrath et al. 1998).  We use 
PHOEBE for modeling single stars due to the fact that the most recent version has 
been adapted for the $Kepler$ mission (with new stellar atmosphere models and limb 
darkening coefficients). We simply set the orbital period of the binary to the 
rotation period, and turn-off the light from the companion star. 

With limited details about our stars, there are few constraints on the
input parameters used in light curve modeling. In the following, we have assumed
that our objects are main sequence dwarfs with the masses, radii, and log$g$ expected 
for stars with the T$_{\rm eff}$ listed in the $KIC$. There are four relevant
parameters for adding a spot to generate a light curve using PHOEBE: the spot
longitude, co-latitude, radius, and temperature (input as the ratio 
T$_{\rm spot}$/T$_{\rm eff}$). Co-latitude has its normal meaning in that it is
the angle between the pole of the star, and the center of the spot. Radius is the 
angular radius of the spot as measured at the center of the star. The minima in 
light curves for stars with a single spot are driven mostly by the interplay between 
the inclination angle and co-latitude, and less by spot size and temperature contrast, 
which are somewhat, but not fully, degenerate. PHOEBE assumes round spots of uniform 
temperature.

To demonstrate the difficulties with ascertaining the
exact spot distributions for any one light curve, we return to K10200948. First,
if we were to classify the Q3 light curve for this object, we would have deemed
it ``continuously variable''. The Q5 light curve, however, shows it to have
flat-maxima for the last four maxima (and a flat minimum for the first part
of the light curve!). Obviously, this type of issue could erupt for every object in 
our sample through the appearance and/or disappearance of a second spot.

In Fig. 11a we present the Q5 light curve for K10200948 phased to its rotation
period. It is clear that the phasing cleans-up most of the light curve, and it
emphasizes the flat-topped nature of its maxima. We found that the best fit,
one spot model (middle panel) for this light curve occurs with an inclination angle of 
$i$ = 70$^{\circ}$, and a co-latitude of $b$ = 45$^{\circ}$. For this model the spot has
a radius of 10$^{\circ}$, and a temperature ratio $w$ = 0.89. At larger co-latitudes
the shoulders on the minima are too square, at smaller co-latitudes the model light
curves are too sinusoidal. There is a set of models with $i$ $\approx$ 50$^{\circ}$ and 
$b$ $\approx$ 60$^{\circ}$ that fit equally well. In both families of models, the spot
transits a similar line-of-sight chord ($i$ + $b$ $\sim$ 110$^{\circ}$). In all
spot models the shoulder on the light curves disappears for $i$ + $b$ $\leq$ 
90$^{\circ}$. As noted above, these are circumpolar spots that never fully disappear 
for the viewer, leading to continuously changing light curves.

In Fig. 11b we present the phased Q3 light curve for K10200948. The rapidly changing
maxima during this quarter results in a messier phased light curve, but shows
relatively symmetric minima and maxima at roughly twice the rotation period. To
model this light curve we added a second spot at a longitude that is located 
175$^{\circ}$ from the spot used to model the Q5 light curve. For the resulting
models we assumed that both spots had identical co-latitudes. Again, the best
fitting model was one with an inclination angle of $i$ = 70$^{\circ}$, and a 
co-latitude of $b$ = 45$^{\circ}$. Both spots had the same radius, 12$^{\circ}$,
slightly larger than found for the one spot model while at the same time having
slightly higher temperature ratios: $w$ = 0.91. Changing spot size affects the
total continuum level. Close inspection of Fig. 8 shows that the peaks of the
maxima in Q5 were slightly larger, and thus the size of the spots needed to be 
increased to lower the overall flux level seen in the maxima of the Q3 light 
curve (note: small changes in the normalization
between the two quarters could be the source of this issue). Meanwhile, the flux level 
of the minima remained similar, so the spots needed to be ``less dark'' (hotter) to 
fit these minima. As with the single spot model, an additional family of models
with $i$ $\sim$ 50$^{\circ}$, $b$ $\sim$ 60$^{\circ}$ fit equally well.

Given the number of parameters that have no apriori constraints, it is likely that
there are other two spot models that might exist that can explain the Q3
light curve of K10200948 equally well, but the near-identical morphologies of 
the minima strongly suggests two spots with similar parameters. It is also difficult to
obtain the observed light curve without employing spots that are centered in opposite
hemispheres (or nearly so). That two spots with similar parameters would form
on opposite hemispheres is surprising, but a similar result was found for the 
host star of CoRoT-2 by Lanza et al. (2009). Unfortunately, we do not have the Q4 data
that would allow us to investigate exactly how the two spot phase transitioned to just
a single spot. To further confuse the issue, there is some evidence in these light 
curves for a third spot that appears to modify some of the maxima seen in the Q3 and
Q5 light curves for K10200948. The issues we have encountered in modeling this object 
could obviously be true for nearly every other object in our sample, and thus
we defer modeling the light curves of additional objects to a future effort. 

As Neff et al. (1995) point out, any photometric modulations due to starspots is
the {\it asymmetric} component of the starspot coverage. Thus, the spots that
lead to the observed modulations could be isolated on an unspotted disk, or just
be the largest features on a disk that is randomly covered by numerous smaller
active regions. We do not yet have a good idea of the fractional coverage of
stellar photospheres, nor the exact temperature(s) one should ascribe to starspots.
Obviously, we need to limit one of these parameters to derive the other. It is not
possible to do this with photometry, but spectroscopic observations have shown
promise in constraining spot temperatures. Using the onset of TiO features in
the red end of the visual spectrum, Neff et al. find that for the RS CVn
binary II Peg, the best fit spot temperature is 3500 K. They also derive a 
``quiet'' photospheric temperature of 4800 K for this star, leading to a temperature
factor for the starspots of $w$ = 0.73. Neff et al. find that this value is consistent
with large sunspots ($w$ = 0.70), and what has been found in other active stars
(0.65 $\leq$ $w$ $\leq$ 0.85).

To get a 2\% change in the light curve of a cool star, we need to employ spots that 
have radii of $r$ $\sim$ 10$^{\circ}$, and temperature factors of $w$ $\sim$ 0.85.
Obviously, to get the same effect with bigger spots will require higher temperature 
factors, while smaller spots have to be cooler. The smallest spot radius we can 
employ for modeling the Q5 data for K10200948 is 6$^{\circ}$, assuming a 
completely black spot. At the opposite extreme, a spot with a temperature factor of 
0.99 needs to be 32$^{\circ}$ in radius to get a proper fit to the minima.
Note that the fit of a model with this enormous of a spot is only slightly poorer than
the solution arrived at using a spot with a radius of 10$^{\circ}$. We could make the 
larger spot model fit equally well by simply increasing its co-latitude so as to 
sharpen the shoulders of its light curve maxima.

Fortunately, with the high precision photometry emanating from both $Kepler$ and 
$CoRoT$, it is becoming possible to directly
measure spot sizes using exoplanet transits. Silva-Valio \& Lanza (2011) found
for the rapidly rotating (P$_{\rm rot}$ = 4.46 d), active G7V host of the
transiting exoplanet CoRoT-2, the average spot radius was $\sim$ 2.6 $\pm$ 
1$^{\circ}$, with the largest spots having radii of about twice this value. It is
interesting that the more slowly rotating (P$_{\rm rot}$ = 23.6 d) and cooler
(G9V) host star for CoRoT-7 appears to have much larger spot groups, with radii
on order of $\sim$ 20$^{\circ}$ (Lanza et al. 2010). 

As shown earlier, the average amplitude of the variations in our sample of stars
is 1.5\%. If we assume single spots with $r$ $\sim$ 10$^{\circ}$ and $w$ = 0.89,
the fractional photospheric coverage of such a spot is 0.76\%. Solanki \& Unruh
(2004) discuss the spot coverage for the Sun, and find that it ranges by a factor
of ten, with a mean near 0.165\%. The sizes of individual sunspots has a lognormal
distribution, with the largest spots covering about 0.01\% of the visible photosphere.
If cool dwarfs have a similar lognormal spot distribution, then the mean of the actual,
fractional spot coverage will be closer to 10\% for our rotating sample of stars.

\section{Discussion and Conclusions}

The original goal of our program was to identify new LMEBs of long period so as to
investigate whether the fundamental parameters for the components in those systems 
might more closely resemble the predictions from stellar models. We have been
successful in this quest by finding a new long period LMEB. Unfortunately, this 
object is quite faint, and this result has now been superseded by 
the findings of Coughlin et al. (2011) where numerous
such objects, all brighter than K6431670, were discovered. Follow-up of those
objects is better suited to investigating whether rapid rotation is playing an
important role in creating the discrepancy in radii between observations and models.
While we do find that activity levels in single stars decrease with increasing
rotation period, it can still remain quite high for rotation periods of 
P$_{\rm rot}$ $\leq$ 25 d, so even some long period binaries may harbor components 
with significant rotation-induced activity levels.

The light curves of low-mass stars exhibit quite a large range in behavior. Prior
to $Kepler$, it was difficult to obtain long duration observations of the 
required precision to explore this variability. At a level of 1\%, the majority 
of these late type stars are variable, though there is a strong trend for the 
cooler stars to show a higher incidence of variability. For example,
we find that less than half of the K0 dwarfs in our sample are variable, while
88\% of stars with T$_{\rm eff}$ $\leq$ 4000 K are variable. Early K dwarfs
thus make excellent targets for exoplanet searches.

Dinissenkov et al. (2010) have modeled the angular momentum transport in solar-like
stars and found that after 4 Gyr, all low-mass stars should rotate with periods
similar to that of the Sun. The mean rotation period that we find, $\langle$ 
P$_{\rm rot}$
$\rangle$ = 32.12 d, is completely consistent with this prediction, especially
given that our sample of periodic variables should have roughly same age and
metallicity as the Sun. If ages could somehow be established for our objects,
it would provide useful insight into the angular momentum loss process in a
region of parameter space that is currently only occupied by the Sun. Obviously,
age determinations for isolated late-type dwarfs are extremely difficult, but it is 
possible to measure the kinematic motions of a large sample of such objects and 
arrive at a statistically useful age vs. rotation rate determination.

Equally interesting is the pursuit of the longer period variables. As Vaughan
et al. (1981) have shown, active regions on solar-like stars seem to be able
to persist for several rotation cycles. Our results bear-out this finding with
several objects having rotation periods in excess of 100 d. Is there an upper
limit to the rotation period of late-type stars, or conversely, the lifetime of
active regions on these objects? Out of our 670 target sample, we found 134 objects
that appeared to have sinusoidal light curves with periods in excess of 90 d.
The amplitudes of these variations were generally similar to the rotational 
sample, though few showed variations in excess of 1.5\%. These objects warrant
further $Kepler$ follow-up to enable the determination of whether these light curves
are consistent with rotation and, if so, what are the upper bounds on the rotation 
rates of late-type dwarfs? Solar active regions rarely remain intact for more than a 
single rotation period. The fact that some objects appear to have much longer-lived
features is intriguing, suggesting that some other factor besides rotation influences 
magnetic activity in solar-like stars.

It remains difficult to extract significant insight into the nature of starspots from 
the broad-band light curves of late-type stars, even those with the high precision 
afforded by $Kepler$. The results we have found for the sizes and relative temperatures
of the spots on these stars are fully consistent with those found by others. Future
investigations into starspot parameters using exoplanet transits will be considerably
more useful, though they will only probe a rather small range in latitude for the
exoplanet host stars.
Our results do strongly suggest, however, that the majority of the stars
in our sample either have polar spots, or they have two spots of similar size and 
temperature that are separated by $\sim$ 180$^{\circ}$ in longitude. This conclusion
is further strengthened by the fact that we have numerous objects with flat minima.
Such objects must have more than one spot to attain such a light curve, though 
it is actually quite hard to produce models with flat minima using round spots. This
probably indicates that the active regions on these stars have complex shapes that 
are quite extended in longitude. 

\acknowledgements Kepler was competitively selected as the tenth Discovery mission. 
Funding for this mission is provided by NASA's Science Mission Directorate. The
authors have been partially supported from NASA grant NNX10AC40G. JLC is supported
through an NSF Graduate Research Fellowship.

\clearpage
\begin{center}
{\bf References}
\end{center}

\begin{flushleft}
Baraffe, I., Chabrier, G., Allard, F., \& Hauschildt, P. H. 1998, A\&A. 337, 403\\
Basri, G., et al. 2011, AJ, 141, 20\\
Basri, G., et al. 2010, ApJ, 713, 155\\
Benko, J. M., et al. 2010, MNRAS, 409, 1585\\
Bessell, M. S. 1991, AJ, 101, 662\\
Bopp, B. W. 1987, ApJ, 317, 781\\
Borucki, W. J., et al. 2010, Science, 327, 977\\
Borucki, W. J., et al. 2011, ApJ, 736, 19\\
Brown, T. M., Latham, D. W., Everett, M . E., \& Esquerdo, G. A., 2011, AJ, 142, 112\\
Christensen-Dalsgaard, J., \& Frandsen, S. 1983, Solar Physics, 82, 469\\
Coughlin, J. L., L$\acute{o}$pez-Morales, M., Harrison, T. E., Ule, N., \& Hoffman, D. 
I.  2011, AJ, 141, 78\\
Covey, K. R., et al. 2007, AJ, 134, 2398\\
Delfosse, X., Beuzit, J., -L., Marchal, L., Bonfils, X., Perrier, C.,
S$\acute{e}$gransan, D., Udry, S., Mayor, M., \& Forveille, T. 2004, ASPC, 
318, 166\\
Denissenkov, P. A., Pinsonneault, M., Terndrup, D. M., \& Newsham, G. 2010,
ApJ, 716, 1269\\
Devor, J., Charbonneau, D., Torres, G., Blake, C. H., White, R. J., Rabus, M.,
O'Donovan, F. T., Mandushev, G., Bakos, G. A., Furesz, G., \& Szentgyorgyi, A. 2008,
ApJ, 687, 1253\\
Dorren, J. D., \& Guinan, E. E. 1982, AJ, 87, 1547\\
Duquenney, A. \& Mayor, M. 1991, A\&A, 248, 485\\
Fischer, D. A., \& Marcy, G. W. 1992, ApJ, 396, 178\\
Garcia, R. A., et al. 2011, MNRAS, 414, L6\\
Gilliland, R. L. 2008, AJ, 136, 566\\
Houk, N., Swift, C. M., Murray, C. A., Penston, M. J., \& Binney, J. J. 1997,
``Proceedings of the ESA Symposium {\it Hipparcos - Venice '97}'', ESA SP-402, 
279\\
Huber, D., et al. 2010, ApJ, 723, 1607\\
Jenkins, J. M., et al. 2010, ApJ, 713, L120\\
Johnson, H. L. 1966, ARA\&A, 4, 193\\
Kallrath, J., Milone, E. E., Terrell, D., \& Young, A. T., 1998, ApJ, 508, 308\\
Knaack, R., Fligge, M., Solanki, S. K., \& Unruh, Y. C. 2001, A\&A, 376, 1080\\
Krishnamurthi, A., Pinsonneault, M. H., Barnes, S., \& Sofia, S. 1997, ApJ,
480, 303\\
Kron, G. E. 1947, PASP, 59, 261\\
Lanza, A. F., et al. 2010, A\&A, 502, 53\\
Lanza,  A. F., et al. 2009, A\&A, 493, 193\\
Lanza, A. F., Rodnono, M., \& Pagano, I. 2004, A\&A, 425, 717\\
Leinert, C., Henry, T., Glindemann, A., \& McCarthy D. W., 1997, A\&A, 325, 159\\
Lenz, P., \& Breger, M. 2005, Comm. in Asteroseismology, 146, 1\\
L$\acute{o}$pez-Morales, M. 2007, APJ, 631, 1120\\
Mayor, M., Duquennoy, A., Halbachs, J., -L., \& Mermilliod, J., -C. 1992, ASPC, 
32, 73.\\
Morales, J. C., Gallardo, J., Ribas, I., Jordi, C., Barraffe, I., \& Chabrier, G.
2010, ApJ, 718, 502\\
Morales, J. C., Ribas, I., \& Jordi, C. 2008, A\&A, 478, 507\\
Murtagh, F., \& Heck, A. 1987, in ``Multivariate Data Analysis'' (Springer-Verlag:
Berlin)\\
Neff, J. E., O'Neal, D., \& Sarr, S. H. 1995, ApJ, 452, 879\\
Ng, Y. K., Bertelli, G., Chiosi, C., \& Bressan, A. 1997, A\&A, 324, 65\\
Pagano, I., Lanza, A. F., \& Rodon$\grave{a}$, M. 2005 in ``Proceedings of the 
13th 
Cambridge Workshop on Cool Stars, Stellar Systems and the Sun'' (ESA: Hamburg), 863\\
Pr$\check{s}$a, A., et al. 2011, AJ, 141, 83\\
Pr$\check{s}$a, A., \& Zwitter, T 2005, ApJ, 628, 426\\
Radick, R. R., Thompson, D. T., Lockwood, G. W., DUncan, D. K., \& Baggett, W.
E. 1987, ApJ, 321, 459\\
Silva-Valio, A., \& Lanza, A. F. 2011, A\&A, 529, A36\\
Solanki, S. K., \& Unruh, Y. C., 2004, MNRAS, 348, 307\\
Southworth, J., Maxted, P. F. L., \& Smalley, B. 2004, MNRAS, 351, 1277\\
Southworth, J., Zucker, S., Maxted, P. F. L, \& Smalley, B. 2004, MNRAS, 355,
986A\\
Vaughan, A. H., Baliunas, S. L., Middelkoop, F., Hartmann, L. W., Mihalas, D., Noyes,
R. W., \& Preston, G. W. 1981, ApJ, 250, 276\\
Verner, G. A, et al. 2011, MNRAS, 415, 3539\\
Zahn, J, 1977, A\&A, 57, 383\\
Zahn, J. 1994, A\&A, 288, 829\\
\end{flushleft}
\clearpage

\clearpage
\begin{deluxetable}{ccccccc}
\renewcommand\thetable{1}
\centering
\tablecolumns{7}
\tablewidth{0pt}
\tablecaption{New Eclipsing Binaries}
\tablehead{
\colhead{Kepler ID} & \colhead{T$_{\rm eff}$}& \colhead{$r$} &\colhead{Period } & \colhead{T$_{\rm 0}$} & 
\colhead{Eclipse Depth} & \colhead{Notes} \\
\colhead{ } & \colhead{(K)} &\colhead{(mag)}  &\colhead{(days)}&\colhead{(JD 2450000+)}&
\colhead{(\%)}&\colhead{} }
\startdata
4636722  & 5119 & 16.24 & 0.4064 & 5004.4722 & 7.9 & 1 \\
6431670  & 5103 & 16.10 & 29.911 & 5097.4786 & 34.4 & 2 \\
7732791  & 4197 & 16.21 & 2.0644 & 5006.1225 & 12.4 & - \\
8086234  & 4160 & 16.52 & 0.2570 & 5093.4422 & 11.1 & 1,3\\
8211824  & 4860 & 16.78 & 0.8412 & 5004.6924 & 3.9 &  3\\
12109845 & 4371 & 16.46 & 0.8660 & 5276.7960 & 9.6 &  4 \\
\enddata
\tablenotetext{1}{This object is probably a W UMa system.}
\tablenotetext{2}{This binary is highly eccentric, with $e$ = 0.33.}
\tablenotetext{3}{This light curve has a significant third light component.}
\tablenotetext{4}{Initial modeling suggests a late K + early M detached binary.}
\end{deluxetable}
\clearpage

\begin{deluxetable}{cccccc}
\tablewidth{0pt}
\renewcommand\thetable{2}
\centering
\tablecaption{Periodic Variables$^{\rm 1}$}
\tablehead{
\colhead{Kepler ID} & \colhead{$r$} &\colhead{T$_{\rm eff}$}&\colhead{P$_{\rm rot}$}
& \colhead{Mean Flux}&\colhead{Amplitude} \\
\colhead{ } & \colhead{(mag)}  &\colhead{(K)}& \colhead{(Days)} & \colhead{(Counts)} & \colhead{(Counts)}}
\startdata
2282506 & 16.345 & 5113 & 68.03 & 3107.3 & 25.0 \\
2835732 & 15.706 & 5197 & 86.21 & 6468.0 & 31.0 \\
2849894 & 16.646 & 3729 & 30.58 & 3413.9 & 85.6 \\
3097797 & 16.608 & 4970 &  8.80 & 2413.2 & 82.8 \\
3216999 & 16.354 & 4929 & 84.74 & 3328.5 & 34.0 \\
3325249 & 16.164 & 4859 & 16.34 & 4699.9 & 82.0 \\
\nodata & \nodata & \nodata  & \nodata & \nodata & \nodata\\
\enddata
\tablenotetext{1}{The full table can be found in the on-line version of the journal.}
\end{deluxetable}
\clearpage

\begin{deluxetable}{cccc}
\tablewidth{0pt}
\renewcommand\thetable{3}
\centering
\tablecaption{Long Period Variables$^{\rm 1}$}
\tablehead{ 
\colhead{Kepler ID} & \colhead{$r$}   & \colhead{K$_{\rm 2MASS}$}    & 
\colhead{T$_{\rm eff}$} \\
\colhead{         } & \colhead{(mag)} & \colhead{(mag)} & \colhead{(K)}}
\startdata
2439966 &16.762 &14.053 &4288 \\
3217079 &16.323 &14.067 &4790 \\
3219572 &15.982 &13.519 &4368 \\
3322804 &16.637 &14.659 &5066 \\
3424248 &16.743 &14.787 &5093 \\
3425772 &16.966 &14.609 &4977 \\
\nodata & \nodata & \nodata  & \nodata\\
\enddata
\tablenotetext{1}{The full table can be found in the on-line version of the journal.}
\end{deluxetable}
\clearpage
\begin{deluxetable}{cccc}
\tablewidth{0pt}
\renewcommand\thetable{4}
\centering
\tablecaption{Aperiodic Variables$^{\rm 1}$}
\tablehead{ 
\colhead{Kepler ID} & \colhead{$r$}   & \colhead{K$_{\rm 2MASS}$}    & 
\colhead{T$_{\rm eff}$} \\
\colhead{         } & \colhead{(mag)} & \colhead{(mag)} & \colhead{(K)}}
\startdata
3530177 &16.590 &14.403 &4684 \\
3958247 &16.342 &13.878 &4376 \\
8087459 &15.801 &13.823 &5044 \\
8289544 &16.753 &14.512 &4625 \\
8458720 &16.336 &14.051 &4475 \\
8678063 &16.717 &14.567 &4738 \\
\nodata & \nodata & \nodata  & \nodata\\
\enddata
\tablenotetext{1}{The full table can be found in the on-line version of the journal.}
\end{deluxetable}
\clearpage

\begin{deluxetable}{cccc}
\tablewidth{0pt}
\renewcommand\thetable{5}
\centering
\tablecaption{Non-Variable Objects$^{\rm 1}$}
\tablehead{ 
\colhead{Kepler ID} & \colhead{$r$}   & \colhead{K$_{\rm 2MASS}$}    & 
\colhead{T$_{\rm eff}$} \\
\colhead{         } & \colhead{(mag)} & \colhead{(mag)} & \colhead{(K)}}
\startdata
3727636 & 16.550 & 14.361 & 4925\\
4141866 & 16.380 & 14.256 & 4971\\
4545268 & 16.024 & 13.961 & 5034\\
5078821 & 16.310 & 13.919 & 4413\\
5607337 & 16.033 & 13.708 & 4304\\
6106815 & 16.873 & 14.769 & 5137\\
\nodata & \nodata & \nodata  & \nodata\\
\enddata
\tablenotetext{1}{The full table can be found in the on-line version of the journal.}
\end{deluxetable}
\clearpage

\begin{deluxetable}{cc}
\renewcommand\thetable{6}
\centering
\tablecolumns{2}
\tablewidth{0pt}
\tablehead{}
\tablecaption{Candidate Exoplanet System K5164255}
\startdata
Kepler mag. & 16.422\\
Inclination & 88.96$\degr$$\pm$0.40\\
Period & 15.37565$\pm$0.00043 d\\
T$_{0}$ (BJD) & 2455006.6084$\pm$0.0014\\
M$_{\star}$  & 0.75$^{\rm 1}$ M$_{\sun}$\\
R$_{\star}$  & 0.728$\pm$0.083 R$_{\sun}$\\
R$_{planet}$ & 0.91$\pm$0.14 R$_{Jupiter}$\\
T$_{planet}$ & 650$^{\rm 2}$ K\\
$a$ & 0.110 AU\\
\enddata
\tablenotetext{1}{Stellar mass based on T$_{\rm eff}$.}
\tablenotetext{2}{Assuming an albedo of 0.3, and a uniform planetary temperature.}
\end{deluxetable}
\clearpage
\begin{figure}
\scriptsize
\epsscale{0.90}
\plotone{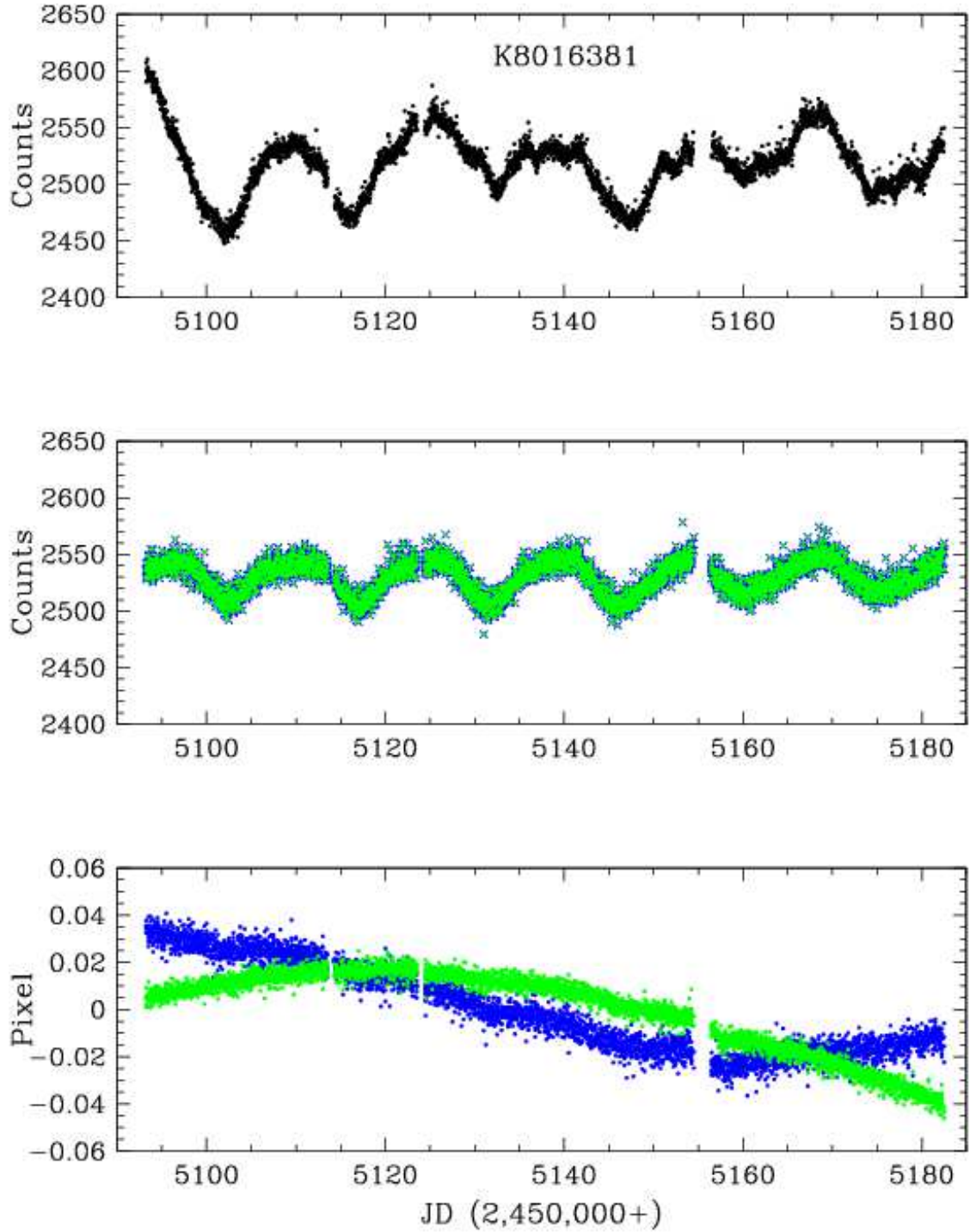}
\caption{a. The $Kepler$ pipeline processed light curve for Kepler ID\#8016381
(top panel), compared to the light curve obtained from the pixel-summed images
(blue crosses, middle panel), and the PCA light curve (green circles, middle
panel) normalized 
to the mean flux of the pixel-summed light curve. These two light curves overlap 
due to the small (and continuous) drift of the centroid of the star within the 
images during Quarter 3 (bottom panel, $x$ in green, $y$ in blue).}
\label{figure1a}
\end{figure}

\renewcommand{\thefigure}{1b}
\begin{figure}
\scriptsize
\epsscale{0.90}
\plotone{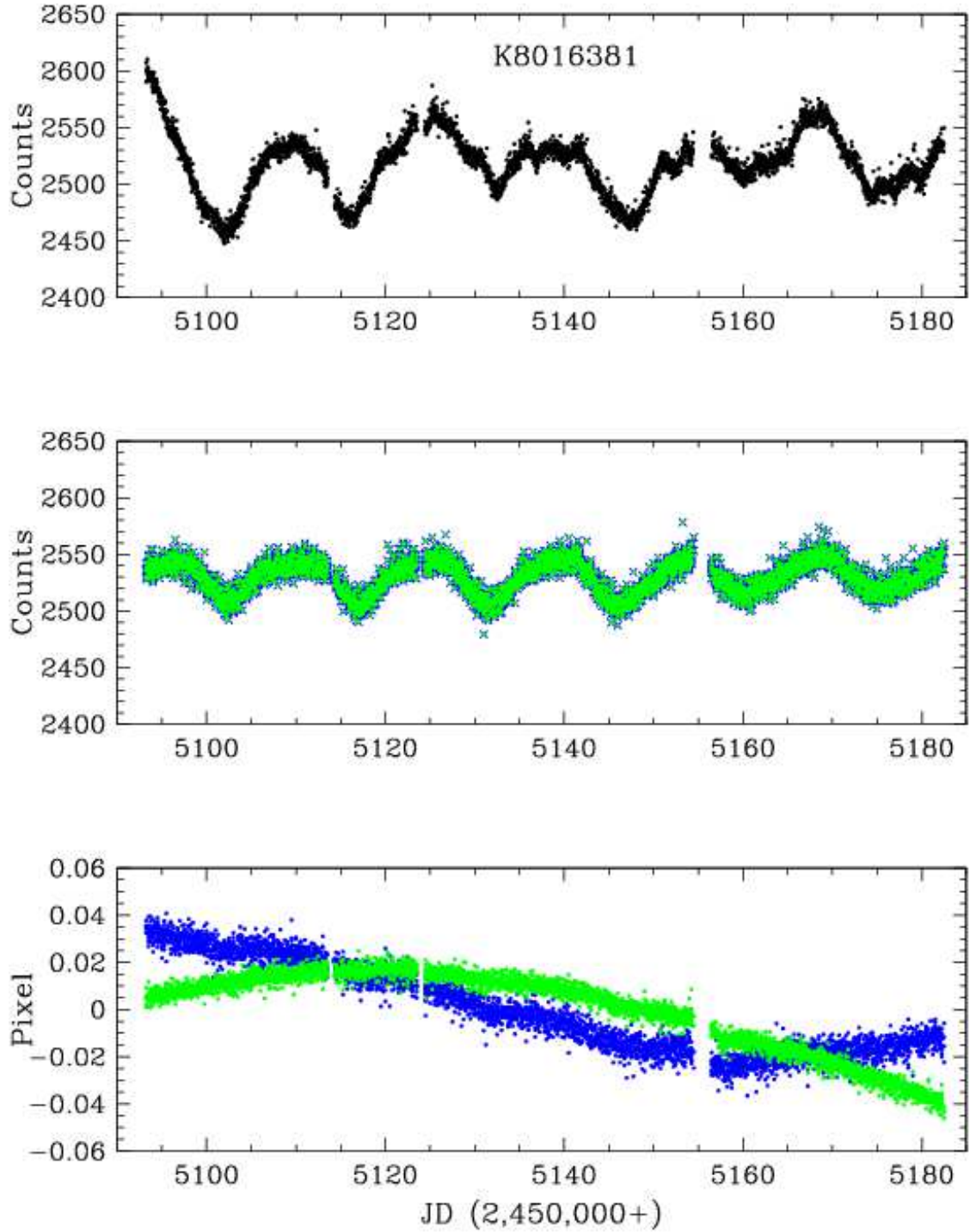}
\caption{The light curves for Kepler ID\#5428432
as in Fig. 1. While our PCA analysis dramatically improved upon the pipeline
processed light curve, it is impossible to discern which variations in the
final light curve are due to intrinsic variability and which are due to the result
of the sudden jumps in the location of the centroid of the star.}
\label{figure1b}
\end{figure}
\renewcommand{\thefigure}{1c}
\begin{figure}
\epsscale{0.90}
\plotone{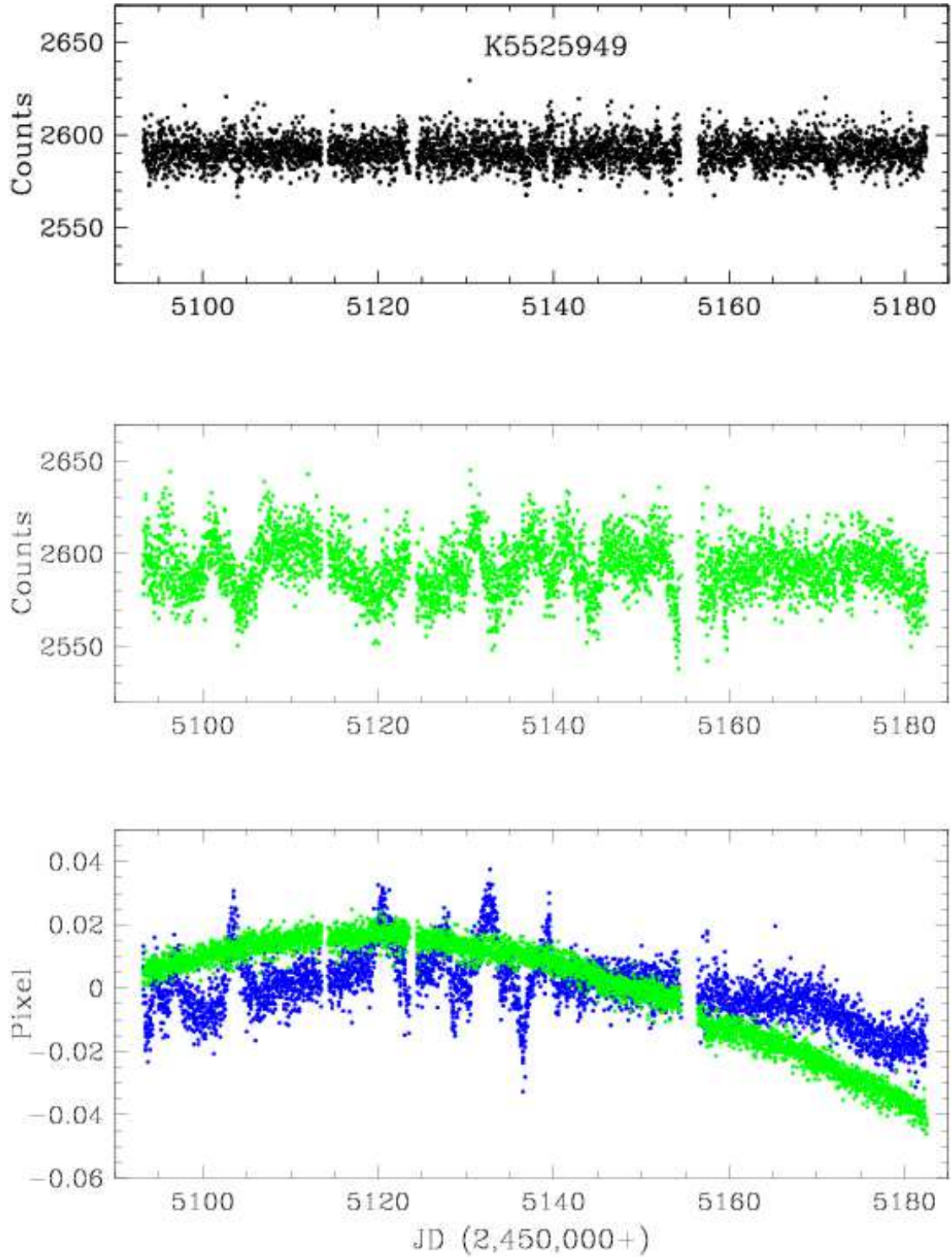}
\caption{The light curves for K5525949. In this case the $Kepler$ light curve
reveals a non-variable source. The PCA light curve reveals an
aperiodic variable due to ``third light contamination'' (the summed-pixel
light curve has not been plotted so as to reduce clutter). As the object (or
background source) brightened/dimmed, the centroid moved in the $-y$/$+y$ 
direction.}
\label{figure1c}
\end{figure}

\renewcommand{\thefigure}{2}
\begin{figure}
\epsscale{1.00}
\plotone{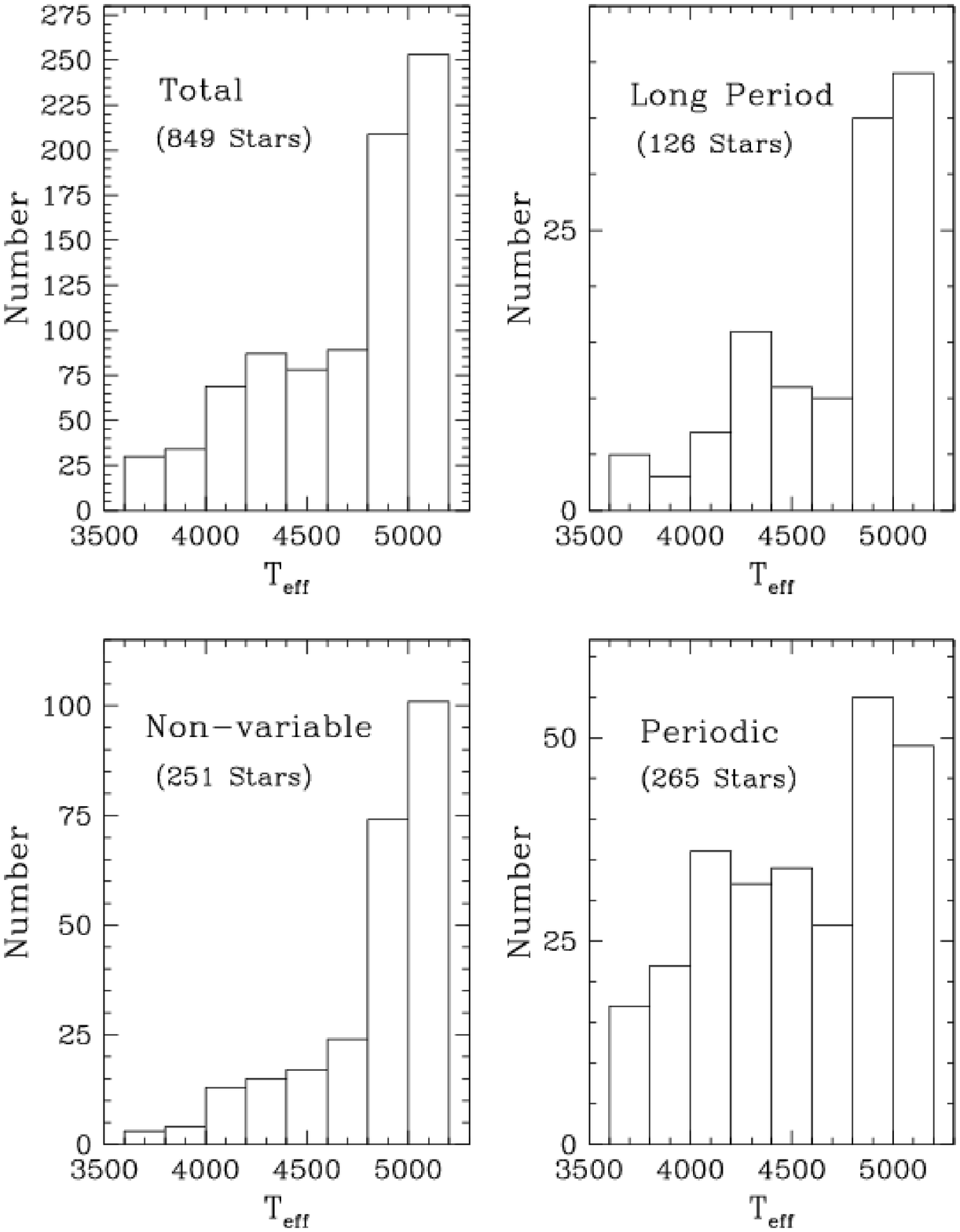}
\caption{A histogram showing the temperature distribution of the entire sample
(top left panel), the non-variable objects (lower left panel),
the long period variables (top right panel), and the periodic
variables (bottom right) sorted into 200 K bins.}
\label{figure2}
\end{figure}

\renewcommand{\thefigure}{3}
\begin{figure}
\epsscale{1.00}
\plotone{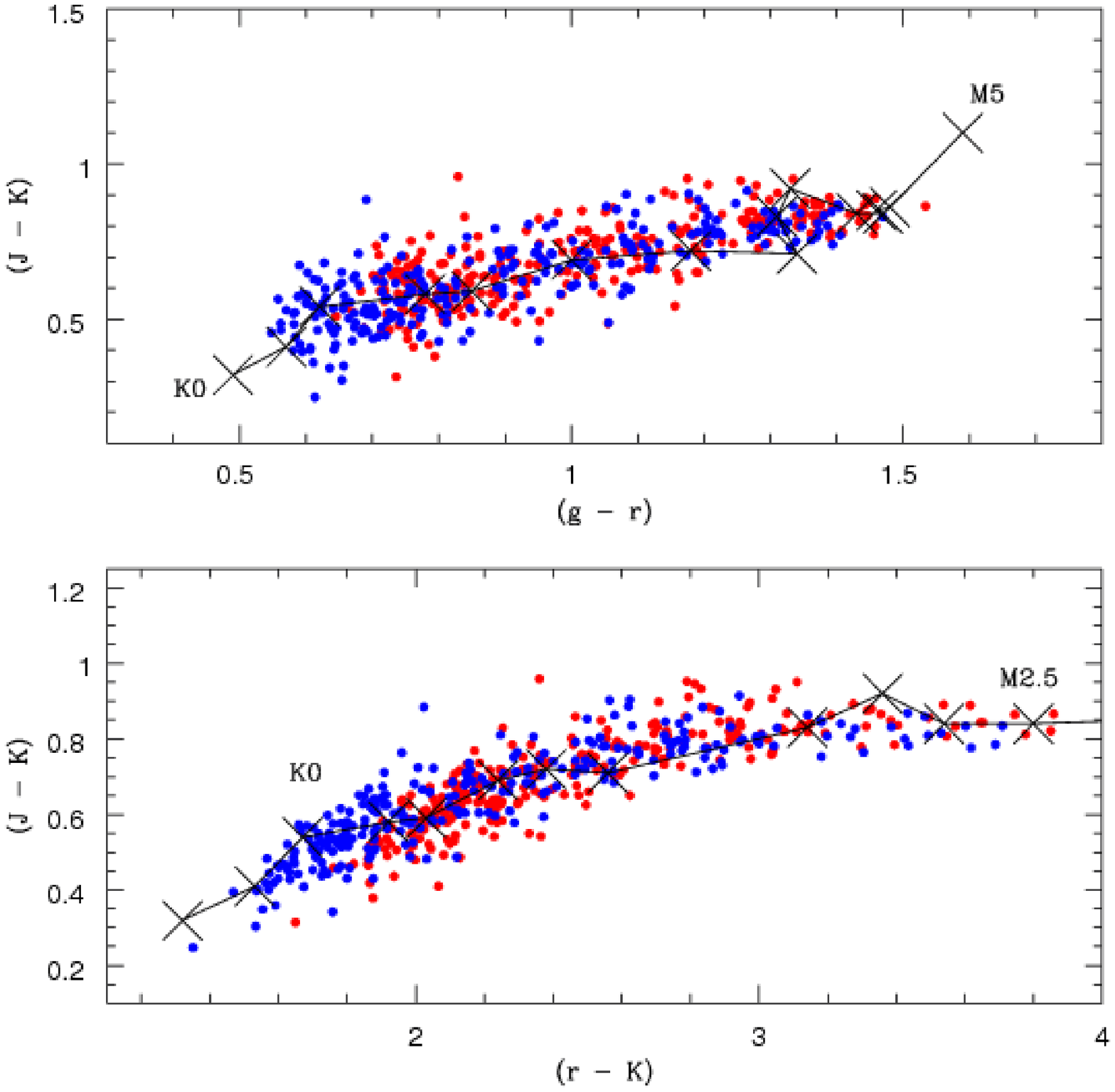}
\caption{Color-color diagrams for the periodic variables in our sample. The
red symbols are the unreddened positions of the objects, while the blue symbols
have been dereddened using the tabulated values of E($B - V$) from the KIC.
In each panel, we have plotted crosses at the locations for main sequence stars 
using the tabulation of synthetic SDSS/2MASS photometry for solar metallicity 
standards from Covey et al. (2007).}
\label{figure3}
\end{figure}

\renewcommand{\thefigure}{4}
\begin{figure}
\epsscale{1.00}
\plotone{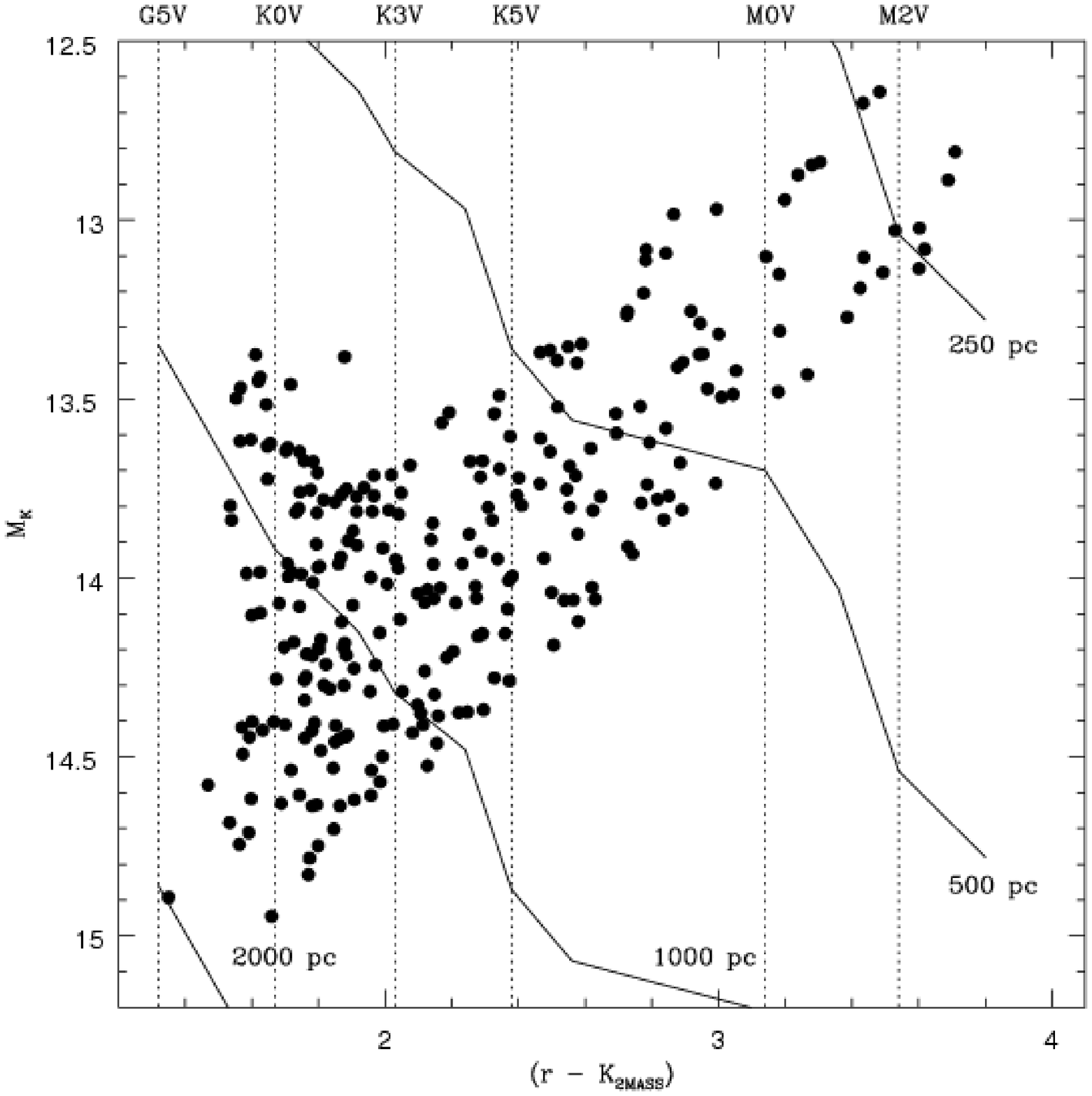}
\caption{An HR diagram for the periodic variables constructed by converting
the ($r$ $-$ $K$) color listed in the KIC into spectral type using Covey et al.
(2007), after accounting for
the tabulated extinction. The absolute visual magnitudes of dwarfs earlier than
K5V have been taken from Houk et al. (1997), and those later than
K7V have been taken from Bessell (1991). The absolute visual magnitudes
were converted to M$_{\rm K}$ using the tabulated ($V$ $-$ $K$) colors of these 
stars in Bessell (1991) and Johnson (1966).} 
\label{figure4}
\end{figure}

\renewcommand{\thefigure}{5}
\begin{figure}
\epsscale{1.00}
\plotone{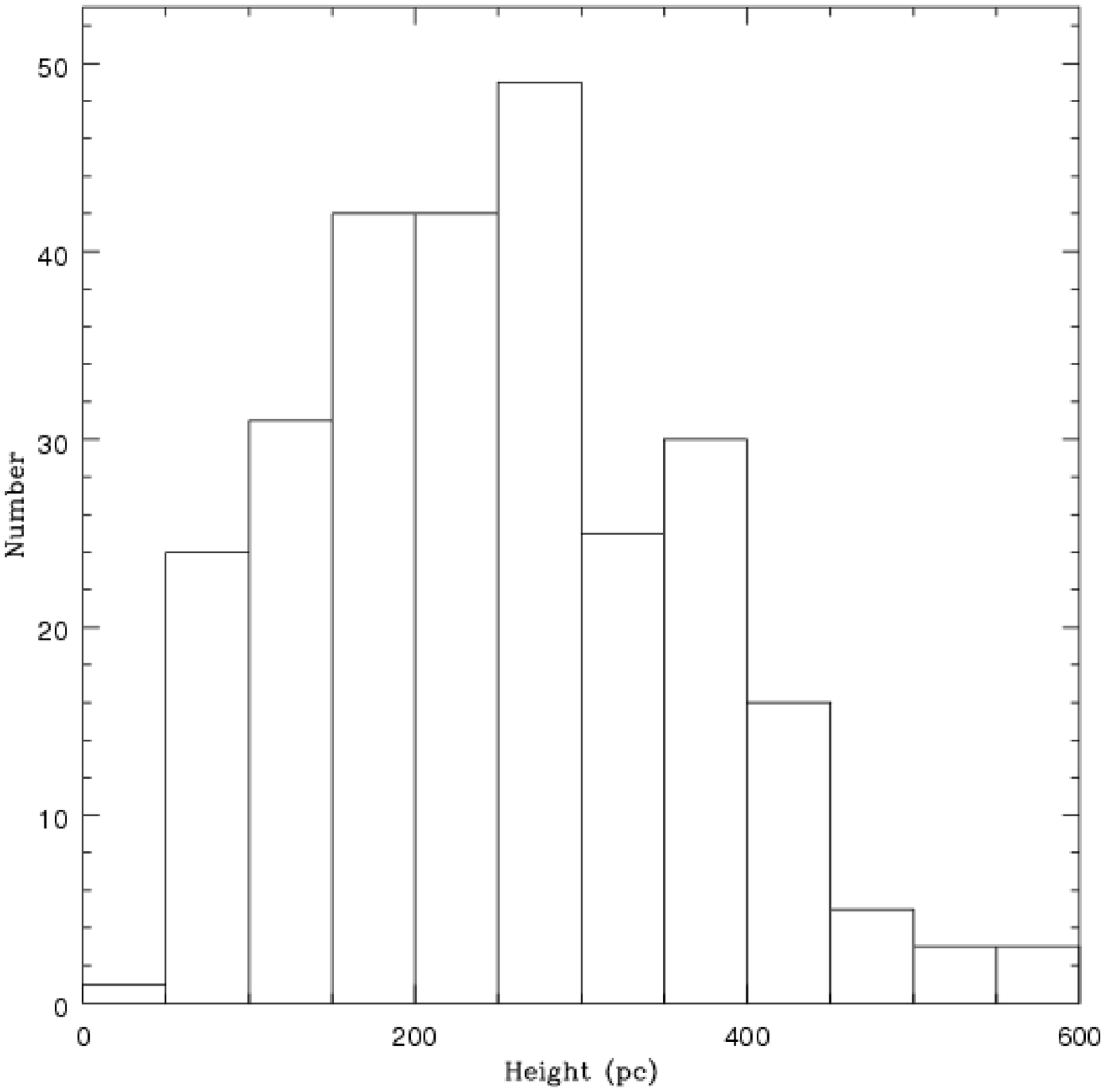}
\caption{A plot of the scale height distribution for the (265) periodic variables
(note that one object has a scale height of 815 pc, and is outside the limits of 
this plot). The mean scale height is 249.9 pc.} 
\label{figure5}
\end{figure}

\renewcommand{\thefigure}{6}
\begin{figure}
\epsscale{1.00}
\plotone{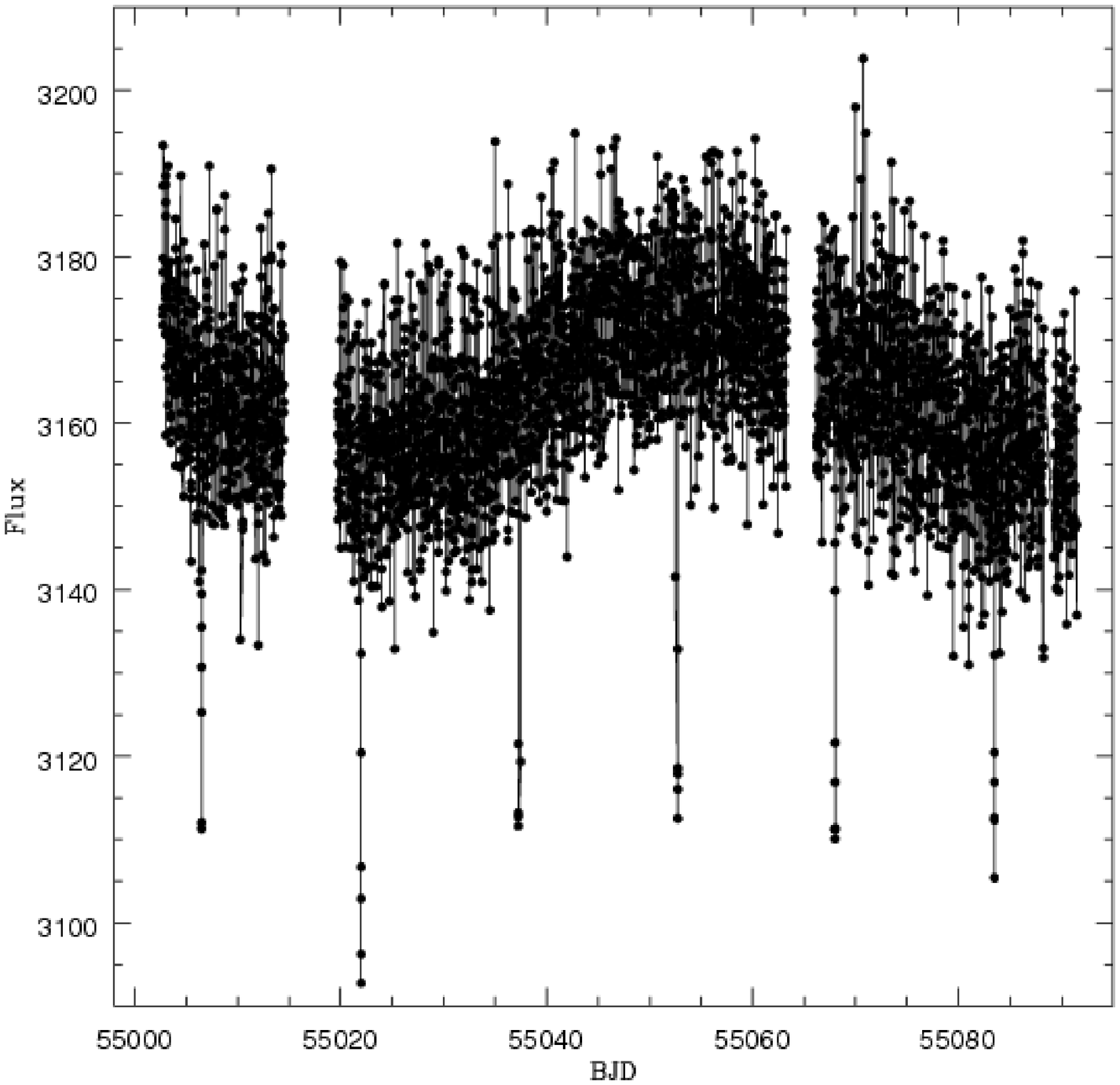}
\caption{The PCA corrected light curve of K6431670, the host of a candidate 
exoplanet that has an orbital period of 15.4 d.}
\label{figure6}
\end{figure}

\renewcommand{\thefigure}{7}
\begin{figure}
\epsscale{1.00}
\plotone{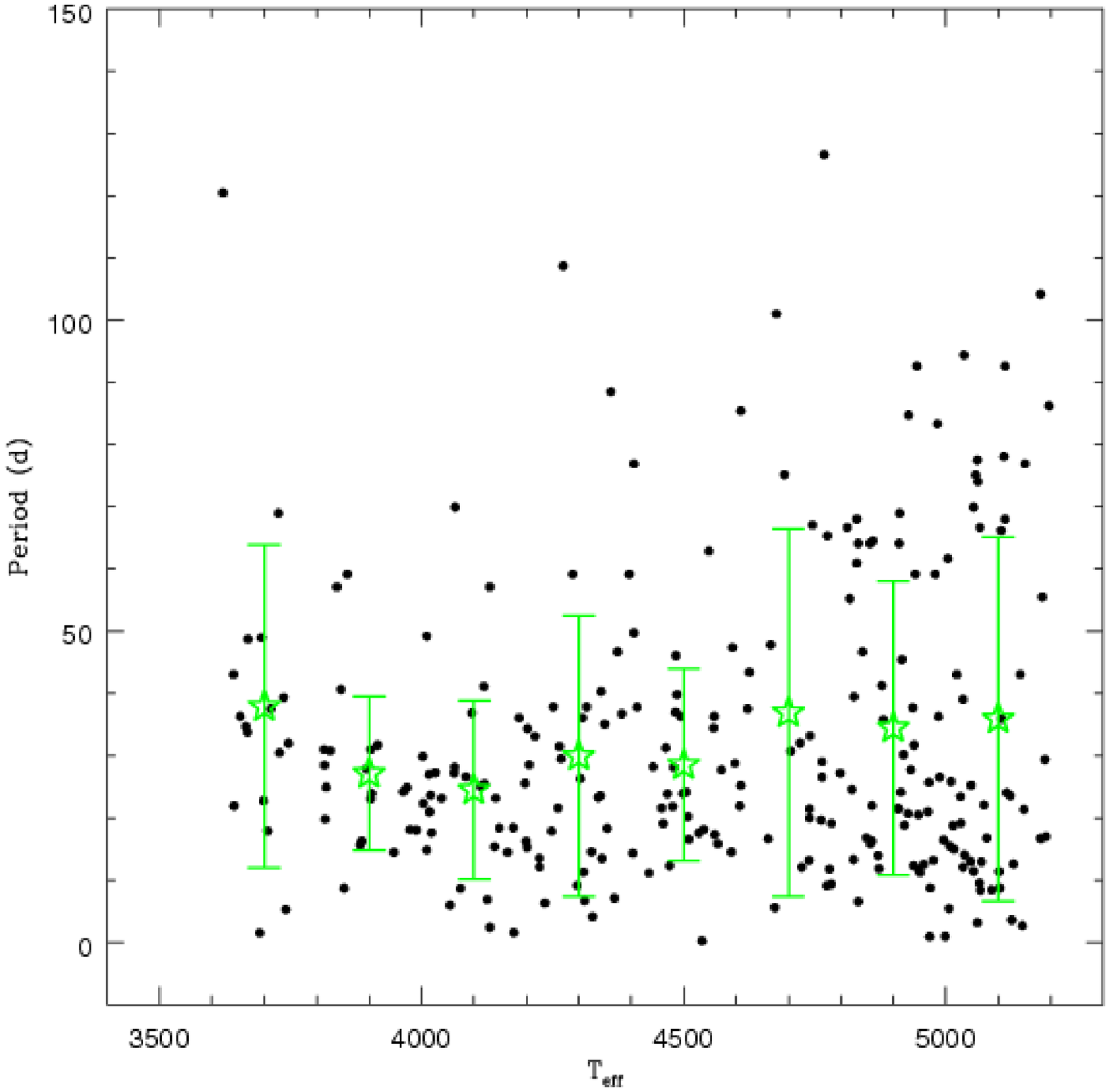}
\caption{The derived rotation periods plotted vs. T$_{\rm eff}$ (black dots). 
The means in 200 K bins are plotted as green stars with error bars.}
\label{figure7}
\end{figure}

\renewcommand{\thefigure}{8}
\begin{figure}
\epsscale{0.80}
\plotone{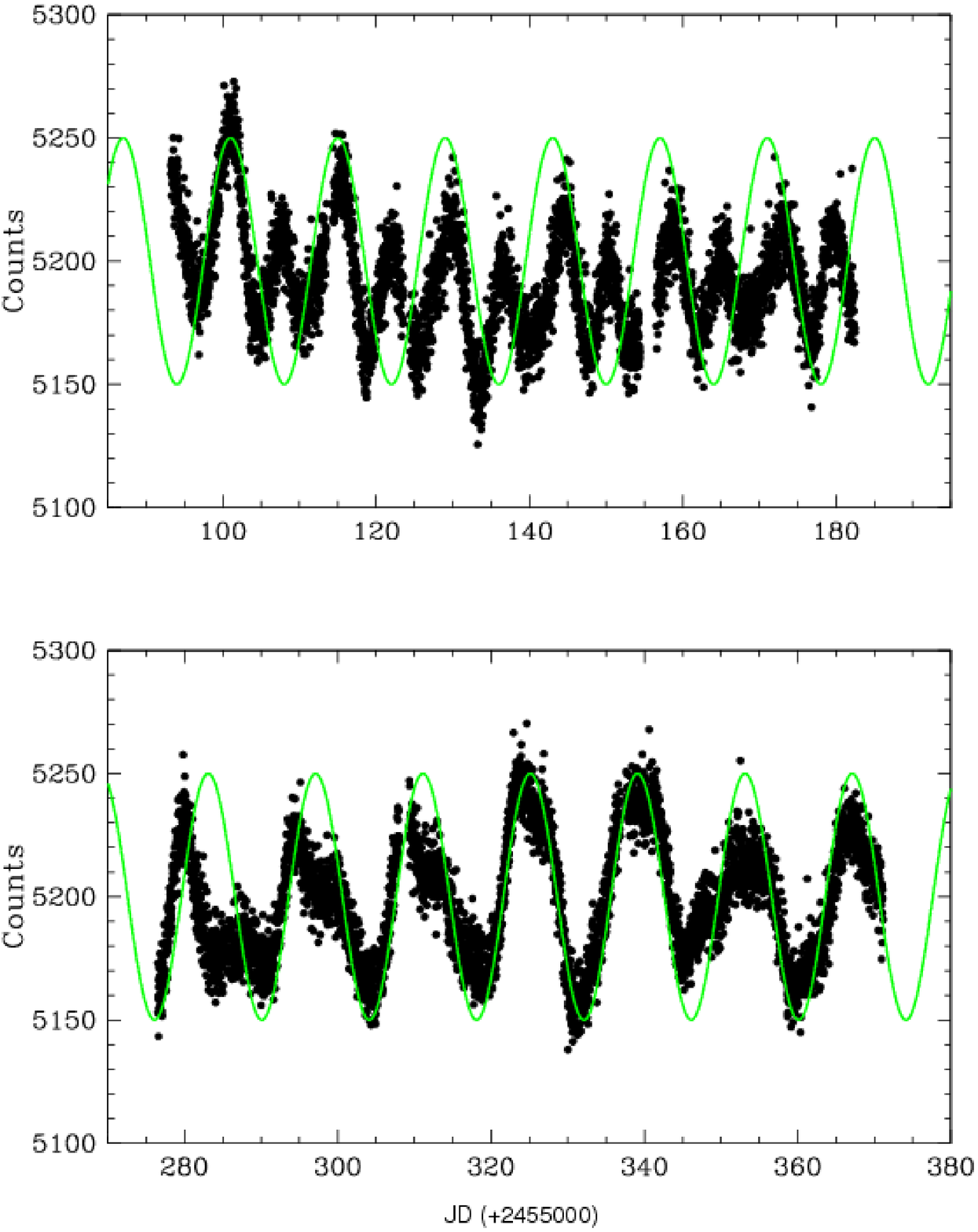}
\caption{A plot of the light curve for K10200948 for Q3 (top panel) and Q5
(bottom panel). The green sine wave plotted in both panels has identical phasing
at the rotation rate tabulated for this object. Clearly this object has one
major spot group that maintains a coherent phasing while one, or more, additional
spot groups is rapidly evolving in size, and/or moving in phase relative to
the dominant group.}
\label{figure8}
\end{figure}

\renewcommand{\thefigure}{9}
\begin{figure}
\epsscale{1.00}
\plotone{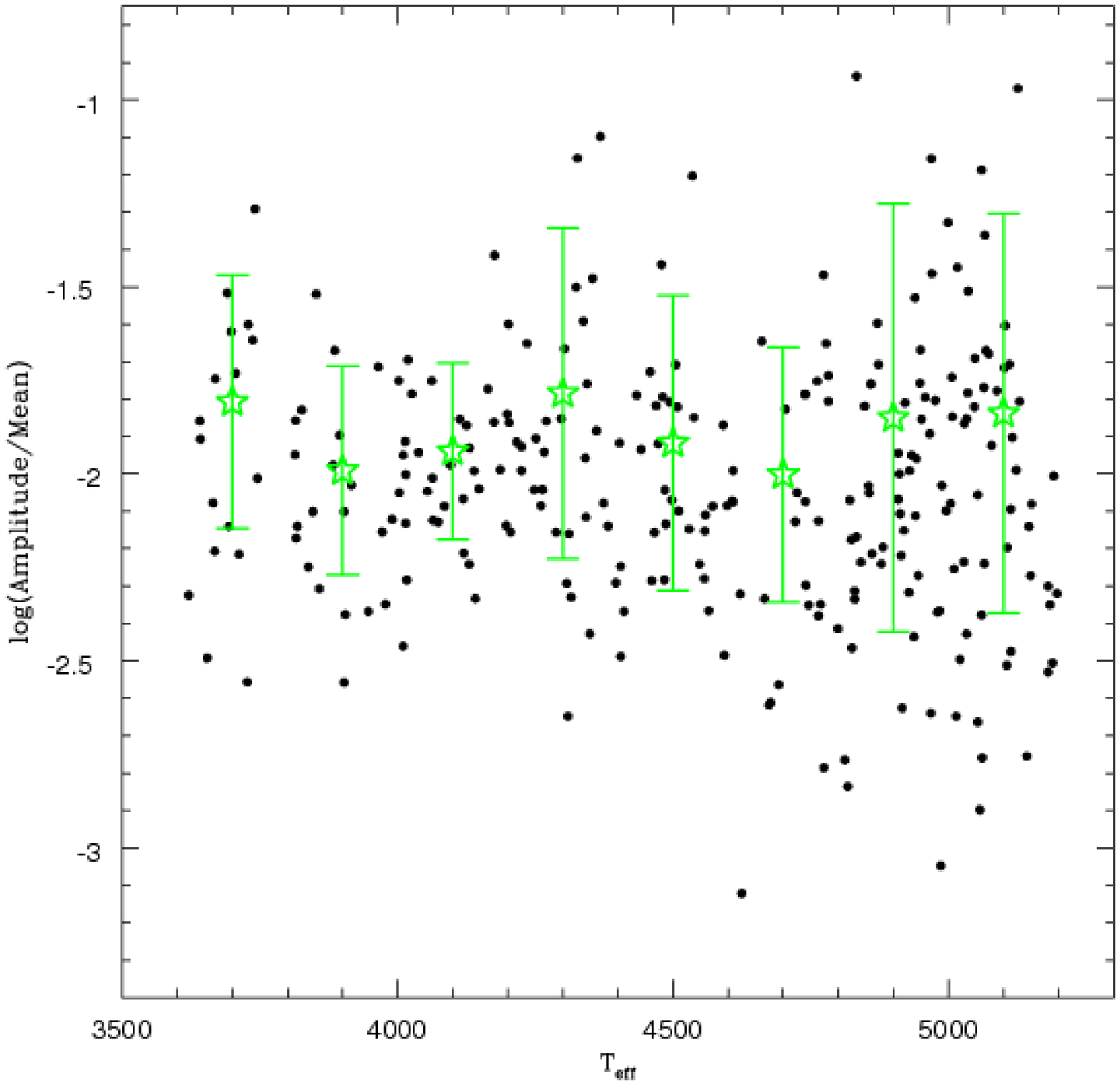}
\caption{The (log) of the ratio of the peak-to-peak amplitude to the light curve
mean plotted vs. T$_{\rm eff}$ for our periodic variables. The means in 200 K bins 
are plotted as green stars with error bars.}
\label{figure9} 
\end{figure}

\renewcommand{\thefigure}{10}
\begin{figure}
\epsscale{1.00}
\plotone{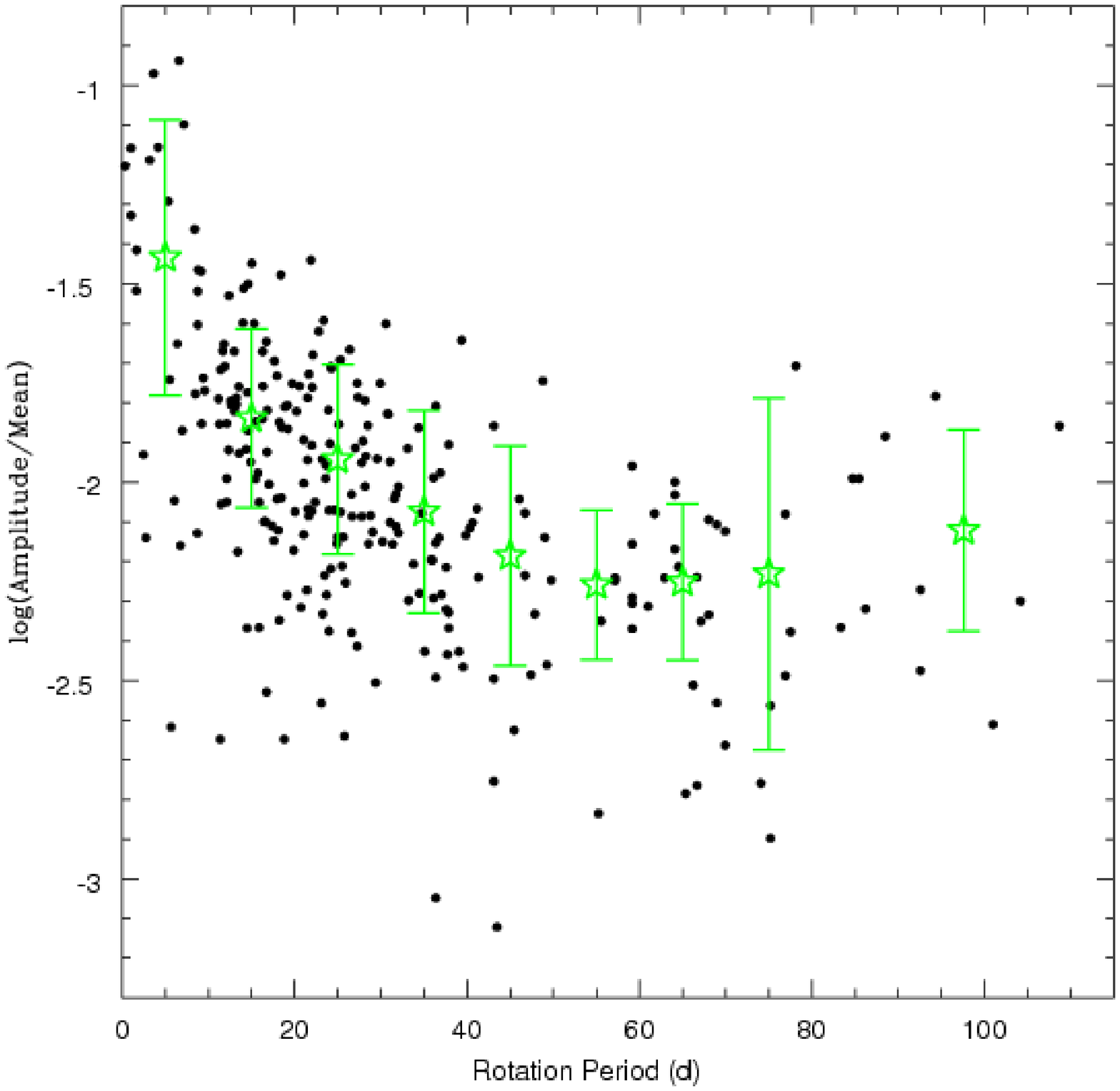}
\caption{The (log) of the peak-to-peak amplitude of the periodic variables vs.
the rotation period. The green stars with error bars are the means in 10 d bins (except
for the final point which contains all objects with rotation periods greater than 80 d).}
\label{figure10}
\end{figure}

\renewcommand{\thefigure}{11a}
\begin{figure}
\epsscale{0.80}
\plotone{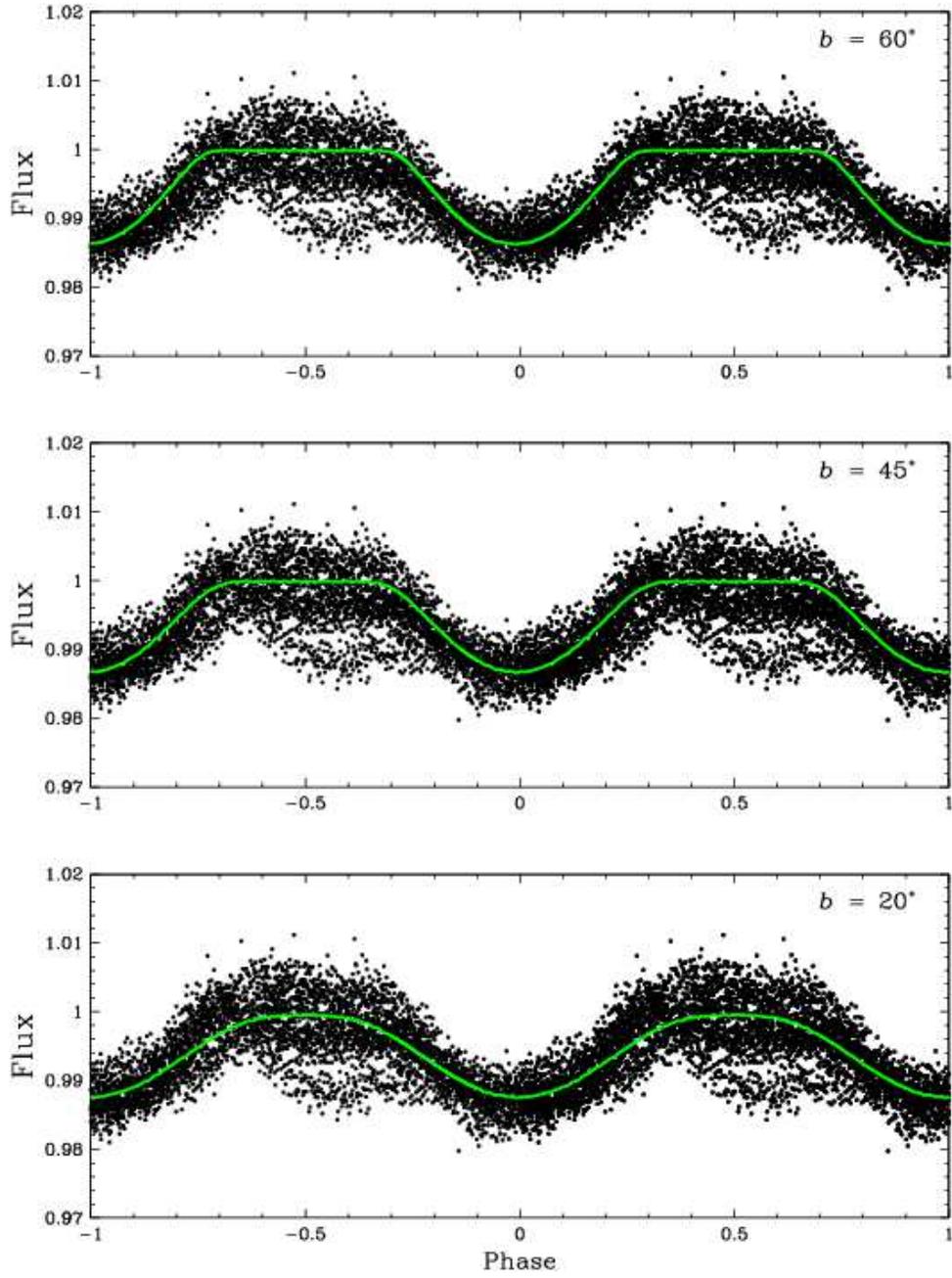}
\caption{The phased Q5 light curve of K10200948. Single spot models (green curves)
with $i$ = 70$^{\circ}$ are plotted in each panel. The co-latitude of the spot,
$b$, is listed in top right corner of each panel.}
\label{figure11a}
\end{figure}

\renewcommand{\thefigure}{11b}
\begin{figure}
\epsscale{0.80}
\plotone{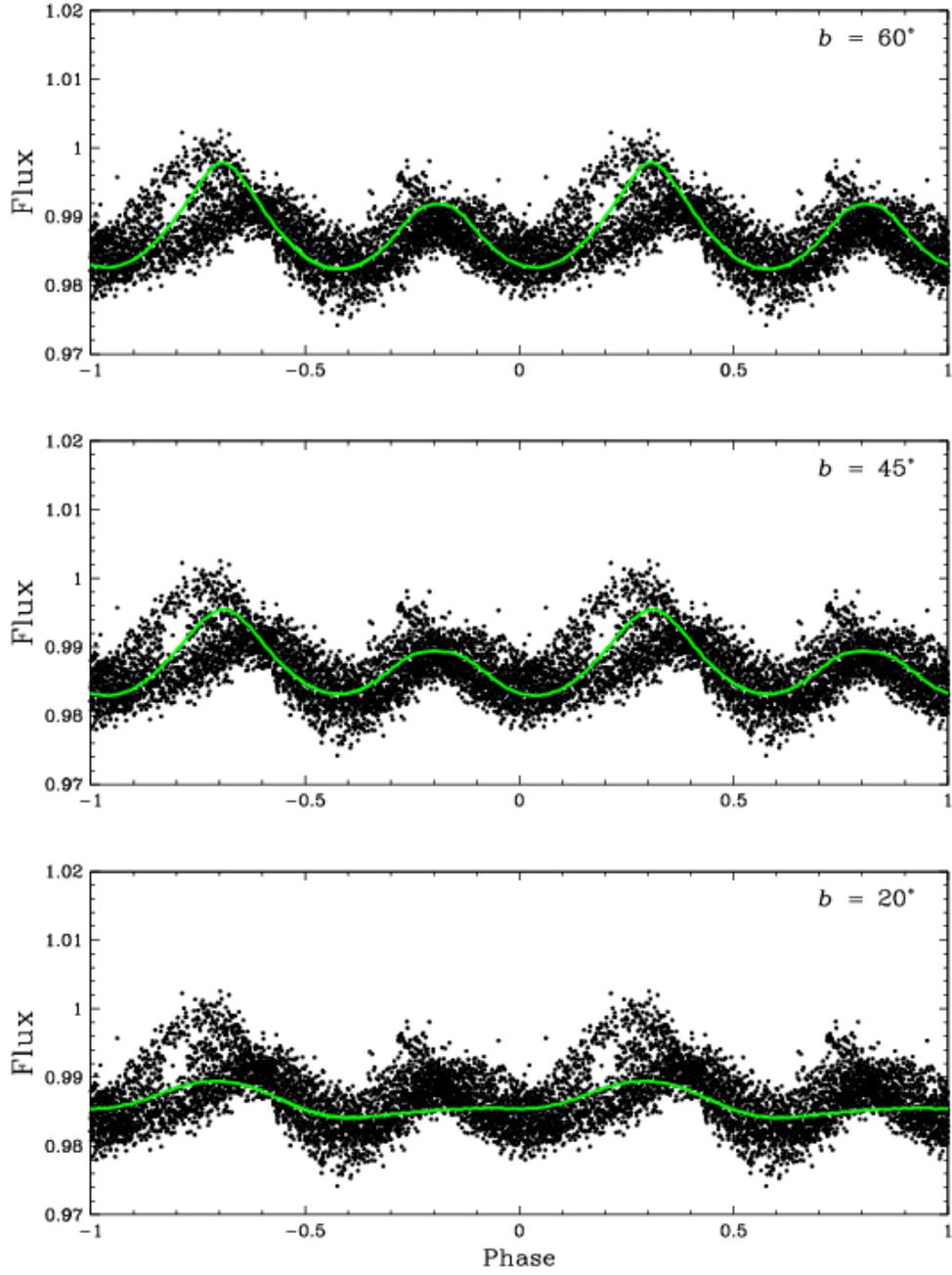}
\caption{The phased Q3 light curve of K10200948 as in Fig. 11a, but for two spot 
models.}
\label{figure11b}
\end{figure}

\end{document}